\documentclass[usenatbib,twocolumn,dvipsnames]{aastex63}
\usepackage{graphicx,amsmath,amssymb,hyperref,xspace}

\hypersetup{
    colorlinks,
    linkcolor={red!60!black},
    citecolor={blue!60!black},
    urlcolor={blue!80!black}
}

\newcommand{\Agama}{\textsc{Agama}\xspace}
\newcommand{\Smile}{\textsc{Smile}\xspace}
\newcommand{\Leiden}{\textsc{Leiden}\xspace}
\newcommand{\Nukers}{\textsc{Nukers}\xspace}
\newcommand{\Monica}{\textsc{MasMod}\xspace}
\newcommand{\Remco} {\textsc{Heidelberg}\xspace}

\renewcommand{\d}{\mathrm{d}}
\newcommand{\D}{\partial}
\newcommand{\Bx}{\boldsymbol{x}}
\newcommand{\Bv}{\boldsymbol{v}}

\begin{document}

\title[Schwarzschild method]
{A new implementation of the Schwarzschild method for constructing \\
observationally-driven dynamical models of galaxies of all morphological types}
\author{Eugene Vasiliev}
\email{eugvas@lpi.ru, mvalluri@umich.edu}
\affil{Institute of Astronomy, University of Cambridge, Madingley road, Cambridge, UK, CB3 0HA}
\affil{Rudolf Peierls Centre for Theoretical Physics, Keble road, Oxford, UK, OX1 3NP}
\affil{Lebedev Physical Institute, Leninsky prospekt 53, Moscow, Russia, 119991}
\author{Monica Valluri}
\affil{Department of Astronomy, University of Michigan, 1085 S. University Avenue, Ann Arbor, MI, USA, 48109}

\begin{abstract}
We present \textsc{Forstand}, a new code for constructing dynamical models of galaxies with the Schwarzschild orbit-superposition method. These models are constrained by line-of-sight kinematic observations and applicable to galaxies of all morphological types, including disks and triaxial rotating bars.
Our implementation has several novel and improved features, is computationally efficient, and made publicly available. Using mock datasets taken from $N$-body simulations, we demonstrate that the pattern speed of a bar can be recovered with an accuracy of $10-20\%$, regardless of orientation, if the 3D shape of the galaxy is known or inferred correctly.\\[5mm]
\end{abstract}

\section{Introduction}  \label{sec:introduction}

The orbit-superposition approach was introduced by Martin \citet{Schwarzschild1979} as a practical method for constructing triaxial stellar systems in dynamical equilibrium, whose existence had been conjectured \citep[e.g.,][]{Binney1978} but not previously demonstrated. In this method, the distribution function (DF) of stars is represented as a weighted superposition of $\delta$-functions in the space of integrals of motion -- in practice, numerically computed orbits in the given potential. Dynamical self-consistency requires that the density generated by this weighted ensemble of orbits is related to the potential via the Poisson equation. To ensure this, the density profile of each orbit $\rho_i(\Bx)$ and of the entire system $\rho(\Bx)$ is discretized into a number of basis elements $m_{ik}, M_k$, and the weights of orbits $w_i$ are assigned in a way that solves the linear equation system $\sum_{i=1}^{N_\mathrm{orb}} w_i\,m_{ik} = M_k$ for all $k$, with the restriction that $w_i \ge 0$.

Many subsequent studies used this method to explore the properties of various stellar systems (e.g., galactic bars, \citealt{Pfenniger1984}, or triaxial galaxies with density cusps, \citealt{Merritt1996}), in particular, the importance of different orbit families, the role of chaos, etc. Another application is generation of equilibrium initial conditions for $N$-body simulations.

It was also quickly realized that this method may be used to construct flexible models of real galaxies, constrained by some sort of kinematic information in addition to the requirement of self-consistency \citep[e.g.,][]{Richstone1984}. In this context, the main focus is on the determination of the range of gravitational potentials in which the DF produces adequate fits to the observed kinematics. In particular, almost all stellar-dynamical estimates of masses of central supermassive black holes (SMBH) are performed with the Schwarzschild method \citep[e.g.,][]{Gebhardt2003, Saglia2016}. Large samples of galaxies have been modeled using this method  to measure stellar mass-to-light ratios and dark matter masses from resolved kinematics \citep{Cappellari2006, Thomas2007, Zhu2018}; the latter study also brought the DF and the orbital structure into focus.

The first aspect (construction and analysis of self-consistent models with the given density profile) is more theoretical, while the second (modelling of real galaxies and constraints on their mass distribution) is an application of the method, which is the main focus of the present paper.

Although there were several early efforts (e.g., the measurement of the black hole mass in M87 by \citealt{Richstone1985}, or the studies of the triaxial Galactic bulge by \citealt{Zhao1996} and \citealt{Hafner2000}), the history of modern observational applications of the Schwarzschild method starts with \citet{Rix1997}, who introduced a practical approach for constructing spherical orbit-superposition models constrained by observed line-of-sight velocity distributions (LOSVD) in the form of Gauss--Hermite (GH) moments. With subsequent generalization to axisymmetry \citep{vdMarel1998,Cretton1999} and various other improvements \citep[e.g.,][]{Krajnovic2005,Cappellari2006}, it came to be known as the \Leiden code. 
Another independent implementation of the axisymmetric Schwarzschild method was presented in \citet{Gebhardt2000, Gebhardt2003, Thomas2004, Siopis2009} and many other studies, and is known as the \Nukers code. \citet{Valluri2004} created a third axisymmetric code \Monica, which is also used until present time. Later, a triaxial generalization of the \Leiden code (in effect, an entirely new one) was developed by \citet{vdBosch2008, vdVen2008}; in absence of an official name, we refer to it as the \Remco code. More recently, spherical and axisymmetric variants of the method specifically tuned for dwarf spheroidal galaxies (dSph) were presented in \citet{Jardel2012}, \citet{Breddels2013}, \citet{Kowalczyk2017} and \citet{Hagen2019}.
For completeness, we mention related approaches for constructing models by a linear superposition of basis elements: finite-size DF blocks in the space of integrals of motion \citep{Merritt1993,Jalali2011,Magorrian2019} or $N$-body particles with adjustable weights, as in the made-to-measure (M2M) method \citep{Syer1996}; the relation between the latter and the classical Schwarzschild method is discussed in \citet{Malvido2015}.

In this paper, we introduce yet another implementation of the Schwarzschild orbit-superposition method, which is designed to be both very general and highly optimized.
The new code \textsc{Forstand} (Flexible orbit superposition toolbox for analyzing dynamical models) is largely based on the techniques used in the ``theoretical'' Schwarzschild modelling code \Smile (\citealt{Vasiliev2013, Vasiliev2015}), but has been almost entirely rewritten from scratch and augmented with the ability to deal with various observational constraints. It is included as part of the publicly available\footnote{\url{http://agama.software}} \Agama library for galaxy modelling \citep{Vasiliev2019}. Some of the preliminary results were presented in \citet{IAU}.

The paper is organized as follows. We describe various technical aspects of the code in Section~\ref{sec:code}, highlighting the differences with other existing implementations of the Schwarzschild method. Then in Section~\ref{sec:tests} we perform various tests on mock/simulated data, and stop here: all observational applications are deferred to forthcoming papers. Section~\ref{sec:discussion} discusses several remaining open questions and wraps up.

\section{Code}  \label{sec:code}

\subsection{Potential representation}  \label{sec:potential}

Any dynamical modelling technique deals with the gravitational potential of a galaxy. Earlier studies typically adopted simple parametric models for the potential (e.g., a flattened logarithmic profile for the dark halo), or represented the density as a Multi-Gaussian Expansion (MGE, \citealt{Emsellem1994, Cappellari2002}), for which the potential can be computed by a 1d numerical quadrature.

Following \citet{Vasiliev2015}, we use an approach where the potential is represented using two very general and flexible approximations: \texttt{Multipole} expansion for spheroidal components (bulge, halo) and/or \texttt{CylSpline} azimuthal-harmonic expansion for disk and bar components. They are described in more detail in Section~2 of \citet{Vasiliev2019} and in the Appendix of \citet{Vasiliev2018}.
The former approach is well-known and used in all four major Schwarzschild codes, but it becomes  inaccurate for strongly flattened systems, for which the latter method is preferable.
Given an arbitrary triaxial%
\footnote{The potential approximations can be used for even more general density profiles lacking triaxial symmetry, but the orbit-superposition technique assumes a steady-state system, presumably excluding non-triaxial features such as spiral arms or lopsided perturbations (although see \citealt{BrownM2013} for a counter-example in the context of the eccentric nuclear disk of M31).}
density profile, the potential is pre-computed to any desired accuracy and stored on an interpolation grid; the subsequent orbit integration uses this interpolated potential and is very efficient (the cost of evaluation is roughly the same for both approaches).
In practice, the density profile can be taken either as a sum of analytic models (S\'ersic, two-power-law, MGE, etc.), or -- for the tests on mock data described in Section~\ref{sec:tests} -- directly from an $N$-body snapshot.

\subsection{Deprojection}  \label{sec:deprojection}

The Schwarzschild method was originally designed to construct dynamically self-consistent models, in which the weighted combination of orbits reproduces the 3D density profile corresponding to the potential in which these orbits were integrated. 
In some cases, for instance, when modelling dwarf galaxies, which are assumed to be dark matter dominated, one may both ignore the contribution of stars to the total potential and skip the self-consistency constraints in modelling, instead only requiring the model to satisfy observable photometric and kinematic constraints. However, in general this is not possible, and one needs to determine the intrinsic 3D density profile from the observed 2d surface brightness profile.

This deprojection problem is a severe obstacle, because already from the dimensional considerations it is clear that the solution is non-unique. In fact, even for axisymmetric systems, the intrinsic density (a function of two coordinates) cannot be uniquely determined, except the edge-on case, as shown in a brief note by \citet{Rybicki1987} and later explored in detail by \citet{Gerhard1996} and  \citet{Kochanek1996}. 
In order to construct a Schwarzschild model, we need to explore systematically the range of 3D density profiles consistent with observations. Unless the galaxy contains a thin gaseous or stellar disk, which can be used to determine the inclination, there is a range of possible inclinations, and for each value there could be a range of possible 3D shapes. \citet{Romanowsky1997} and \citet{Magorrian1999} present practical algorithms for constructing a series of smooth, regularized solutions for the intrinsic density profile in the axisymmetric case, and \citet{Chakrabarty2010} proposed an even more general Bayesian deprojection approach. Some image fitting programs such as \textsc{Imfit} \citep{Erwin2015} can operate with families of parametric 3D density profiles, which are compared to the surface brightness maps after integrating along the line of sight.

On the other hand, if the 3D density follows an ellipsoidally-stratified profile (i.e., equidensity surfaces are concentric ellipsoids with constant axis ratios), then its projection is also stratified on concentric ellipsoids, with the relations between intrinsic and projected axis ratios and viewing angles given, e.g., by \citet{Binney1985} or \citet[Section~3]{vdBosch2008}. Under this assumption, the observed surface brighthess profile composed of one or several ellipsoidally-stratified components (e.g., an MGE) can be deprojected uniquely for a given orientation (except some degenerate cases). This is the approach taken by the vast majority of papers in the literature, and it appears to produce reasonable results for elliptical galaxies. However, bars are manifestly \textit{not} ellipsoidal (most often, boxy) in shape, and the biases arising from incorrect assumptions on the intrinsic shape are poorly known. Figure~2 in \citet{IAU} illustrates that even in the axisymmetric case, the deprojection of an MGE fit to a disk galaxy seen at an intermediate orientation produces a substantially different 3D density profile from the true one, and this biases the measurement of the BH mass.

In the present paper, we do not address the deprojection problem, rather we test the method on the mock data generated from $N$-body simulations, for which the 3D shape is known. We defer a detailed treatment of deprojection of the light distribution and its application to observed galaxies to a later study.

\subsection{From light to mass}  \label{sec:ML}

Even assuming that the intrinsic light density profile could be determined, the mass density profile needs to be specified. Most often, a constant (but \textit{a priori} unknown) mass-to-light ratio $\Upsilon$ is assumed for the entire stellar population, which is then constrained by kinematics. Several studies have explored the effect of a variable stellar M/L (e.g., \citealt{McConnell2013} allowed for a radial gradient of $\Upsilon$, and \citealt{Erwin2018} used two different values for the bulge and the disk components). When using an MGE representation of the density profile, one may ascribe different values of $\Upsilon$ to different Gaussian components, approximating the radial variation inferred from stellar population modelling \citep[e.g.,][]{Valluri2005,Nguyen2018}.

An often-used trick is to rescale all mass components in the galaxy (central SMBH, dark halo, etc.) by the same factor $\Upsilon$, retaining the self-similarity of the potential. In this case, a series of rescaled mass models has the same orbital structure, but the values of velocity recorded in the orbit library should be multiplied by $\sqrt{\Upsilon}$ before comparing to the observations. Therefore, the same orbit library can be reused multiple times, but the optimization problem needs to be solved separately for each $\Upsilon$.

\subsection{Construction of an orbit library}  \label{sec:orbit_library}

The Schwarzschild method operates in two stages.
First, for a given choice of potential, a large number of orbits $N_\mathrm{orb}$ spanning the entire model are integrated for a sufficiently long time (typically $\mathcal O(10^2)$ orbital periods), and their properties are recorded in a suitable format. Second, the optimization problem is solved to assign the orbit weights in a way that satisfies the constraints as closely as possible. In this second step, only some fraction of orbits receive positive weights, but if this number is too small, the model will be implausibly ``jagged''. Hence, even though the method is intrinsically adaptive, it works best if the orbit library was constructed wisely. On the one hand, it needs to have a large enough variety of orbits to choose from, but on the other hand, it should be tailored to the expected orbital configuration of the stellar system. For instance, in a disk galaxy, one would expect to find most stars on close-to-circular orbits, with $v_\phi \gg v_{R,z}$, hence the initial conditions for the orbits should reflect this anisotropy.

Traditionally, the initial conditions are assigned on a grid designed to sample the entire space of integrals of motion in a regular way. One of these integrals is the energy $E$, with typically $20-40$ bins across the entire model. In axisymmetric systems, the other classical integral is the $z$-component of the angular momentum $L_z$, ranging from 0 to the maximum possible value of a circular orbit $L_\mathrm{circ}(E)$, and for the given $E$ and $L_z$, the non-classical third integral (if it exists) determines the thickness of the orbit in radius, or alternatively, its maximum extent in $z$. 
Schemes for sampling the start space of axisymmetric systems are largely similar between studies. \citet[Figure~3]{Cretton1999} use a regular grid in $E$ and $L_z$, and assign starting points for the given $E,L_z$ at regularly spaced locations on the zero-velocity curve in the meridional plane.
Subsequently, the \Leiden code shifted to sampling the position rather than $L_z,I_3$ on a regular 2d grid in the meridional plane (Figure~6 in \citealt{Cappellari2006}). A similar approach is adopted in the \Monica code \citep{Valluri2004}. The \Nukers code additionally employs a Voronoi tesselation scheme for the surface of section $r$ vs. $v_r$ to avoid repeated sampling of the same phase-space region \citep{Thomas2004}.
In a triaxial system, the start space is typically split into two parts: stationary (dropping orbits from the equipotential surface with zero velocity) produces mostly box and high-order resonant orbits, and principal-plane is similar to the axisymmetric case and produces mostly tube orbits \citep{Schwarzschild1979, Merritt1996, vdBosch2008}.

However, the regular grid-like structure of the start space may lead to artifacts in the resulting orbital superposition. \citet{Vasiliev2012} found that such models also are not in perfect equilibrium when evolved as an $N$-body system, because the integrals (most notably, energy) are sampled only at discrete values, and unavoidable two-body relaxation leads to blurring of the DF and associated changes in the density profile.
Therefore, we use an alternative approach, where the initial conditions are sampled randomly rather than regularly. The position is always sampled from the intrinsic density profile of the given galaxy component (disk, halo, etc.), and the velocity is assigned using one of the two possible methods. The first one is more suitable for spheroidal systems: we construct sphericalized density and potential profiles by averaging the actual ones over the two angles, and then determine the self-consistent, possibly anisotropic DF using the \citet{Cuddeford1991} inversion formula, which generalizes the Eddington and Osipkov--Merritt inversion techniques. In this method, the velocity distribution at a fixed position is the same in $\theta$ and $\phi$, but possibly different in $r$. The second approach is more suitable for disks, and is based on solving the anisotropic Jeans equation for the axisymmetrized potential and density, in the formulation of \citet{Cappellari2008}, but for an arbitrary profile (not necessarily an MGE). The velocity is then drawn from a Gaussian distribution with the computed dispersions, which are different in $R$ and $z$ directions, and a nonzero mean in the $\phi$ direction. Hence it creates an orbit library with a preferred rotation direction. 

Each orbit is integrated typically for 100--200 dynamical times (defined as the period of a circular orbit with the given energy in the equatorial plane of the axisymmetrized potential). We use a slightly modified version of the 8th order Runge--Kutta method \textsc{dop853} from \citet{Hairer1993}, which allows one to obtain high-accuracy interpolated solution at any moment of time regardless of the internal timestep of the integrator. 
When constructing models of barred disk galaxies, the potential is assumed to be stationary in the rotating frame, and the pattern speed $\Omega$ becomes another free parameter in the model. The orbit integration in the rotating frame is only slightly more complicated than in a non-rotating system, and the kinematic observables are recorded in the inertial frame. It is important to keep in mind that figure rotation breaks the equivalence between prograde and retrograde orbits (which otherwise look the same except for the flipped sign of velocity). In any case, the randomly sampled initial conditions do not impose any symmetry between these orbits.

In a general rotating triaxial system, there is an overall symmetry w.r.t.\ the reflection about the equatorial plane (flipping of the sign of both $z$ and $v_z$), although individual orbits in certain families (e.g., banana or saucer orbits) need not be symmetric. The simultaneous change of sign of all three coordinates and velocities also preserves the symmetry of the entire system. Therefore, we impose a fourfold discrete symmetry of each orbit when computing its contribution to the kinematic datacube (in a non-rotating system, this would have been an eightfold symmetry of reflection about each of the three principal planes). For axisymmetric potentials, we further randomize the azimuthal angle $\phi$ before computing the projection of each point, and for spherical potentials we randomly choose the orientation the orbit on the sphere specified by two angles ($\theta$ and $\phi$).

During the orbit integration, we store various associated datacubes which are later used in the modelling: the linear superposition of datacubes of individual orbits is required to match the target constraints as closely as possible. 
The most important targets are the intrinsic density distribution and the line-of-sight velocity distributions on the image plane, considered in the following sections. 
Additionally, we store the 6d samples drawn from each trajectory at random times, which can be used to generate an $N$-body representation of the orbit library (if needed).

To improve the smoothness of orbit-superposition models, \Leiden and \Remco codes use a ``dithering'' approach, in which an individual orbit is split into a bunch of $\sim100$ orbits with nearby initial conditions, and all observable datacubes are averaged over this bunch. It would be straightforward to do this in our code as well; however, we prefer to use a larger number of orbits together with regularization constraints (Section~\ref{sec:regularization}) to achieve this goal.

\subsection{Self-consistency constraints}  \label{sec:density_constraints}

\begin{figure*}
\includegraphics{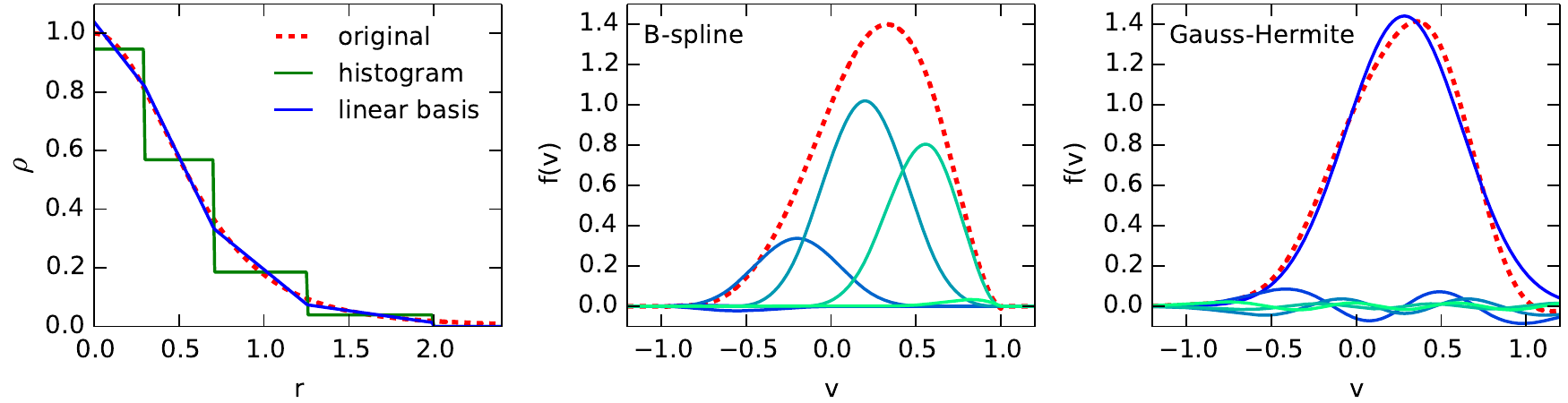}
\caption{
\textit{Left panel}: density discretization using 0th-degree B-splines (histogram, green) or 1st-degree B-splines (piecewise-linear function, blue); the latter approximates the original density (red dotted line) far better and provides additional smoothness constraints in the model. \protect\\
\textit{Center panel}: representation of LOSVD in terms of B-splines of degree 3: contributions of individual basis functions are shown by different shades of blue and green, and their sum by a red dotted line. \protect\\
\textit{Right panel}: representation of LOSVD in terms of Gauss--Hermite series: blue is the dominant term (Gaussian), other colors are higher-order terms starting from $h_3$, and red dotted line is their sum.
}  \label{fig:bsplines}
\end{figure*}

If the model is designed to be dynamically self-consistent (i.e., when stars contribute to the total potential), we need to record the density generated by each orbit, and ensure that it matches the target density of the entire stellar system (or one of its components).

There are several variants of density discretization schemes discussed in the literature. For spherical systems, it is sufficient to store the mass in spherical shells. For axisymmetric systems, the density is discretized on a 2D grid in the meridional plane, typically aligned with spherical coordinates (i.e., radial shells further divided into angular bins). For triaxial systems, the grid is further extended in the $\phi$ dimension \citep[e.g.,][]{vdBosch2008}, or an alternative partitioning scheme with each radial shell divided into three equal segments (in one of the 8 identical octants), and then further into several nearly-equal-area bins (\citealt{Schwarzschild1979, Merritt1996}, see Figure~7 in \citealt{Vasiliev2013}).

A deficiency shared by all these schemes is that they only constrain the average mass in each spatial bin, but provide no control of the smoothness of the mass distribution in a bin. In terms of approximation theory, the function (density) is represented in a discrete way by a histogram, or a basis set with non-overlapping $\sqcap$-shaped basis elements.
However, one may do better by generalizing this scheme to higher-degree finite-element basis sets, as suggested in \cite{Jalali2011}. In common with other parts of the code, we choose B-splines of degree $D$ as the basis set \citep[e.g.,][Chapter IX]{deBoor1978}. Histograms are just B-splines of degree zero, and a better alternative are first-degree B-splines, or $\wedge$-shaped functions spanning two adjacent grid cells (Figure~\ref{fig:bsplines}, left panel).

In the present code, we provide several options for density discretization: cylindrically-aligned meridional-plane grid with the $\phi$ dimension represented by Fourier harmonics (only needed for non-axisymmetric systems, otherwise a single term is used), the scheme of \citet{Merritt1996} for triaxial systems, and a scheme based on multipole expansion of the density \citep{Vasiliev2013}.
The first two options can be used with either the traditional 0th-degree B-splines (histograms), or (preferrably) with 1st-degree B-splines, which provide a better approximation to the target density profile, and additionally enforce smoothness of the density of the orbit-superposition model. In the third scheme, the radial variation of the density is represented as a 1st-degree B-spline, and the angular variation at each radius -- by a spherical-harmonic basis set.
In all variants, the target density profile $\rho(\Bx)$ and the density generated by each orbit $\rho_i(\Bx)$ are discretized in exactly the same way, by computing their Galerkin projections $m_{ik} \equiv \int \rho_i(\Bx)\,B_k(\Bx)\; \d^3\Bx$ onto each element $B_k(\Bx)$ of the basis set. In the traditional discretization scheme, these values are just the cell masses.

\subsection{Kinematics}  \label{sec:kinematics}

Almost all existing Schwarzschild modelling codes are designed to deal only with line-of-sight velocity distribution  functions (LOSVD), not individual stellar velocities, nor proper motions. 
For the Local Group objects, these LOSVD or their moments are constructed by binning up individual stellar velocities \citep[e.g.,][]{Jardel2013,Breddels2013,Kowalczyk2017}, while for most extragalactic objects they come from long-slit or integrated field unit (IFU) spectroscopy. In the latter case, the LOSVDs are measured in some patches (apertures) on the image plane, which may consist of individual spaxels or groups of them, often constructed with the Voronoi binning approach of \citet{Cappellari2003}, and represent the intrinsic LOSVDs convolved spatially with the instrumental point-spread function (PSF). 
Therefore, to take into account the limited spatial resolution, the LOSVD of the model must also be convolved with the instrumental PSF before comparing with the observations, especially when modelling central regions of galaxies around SMBHs, whose radius of influence is often comparable to or smaller than the PSF width.

By contrast, the velocity dimension usually represents the intrinsic (deconvolved) velocity distribution, which comes out of spectral fitting procedure. There are several approaches for deriving the LOSVD from a spectrum in a single spatial bin, and they produce the data in different representations.
In the simplest case, only the mean velocity $\overline v$ and its dispersion $\sigma$ are fitted. This would completely describe the LOSVD if it were a Gaussian, however, in many cases the profiles are strongly non-Gaussian (for instance, asymmetric in the presence of fast rotation, flat-topped or spiky respectively in the cases of tangential or radial anisotropy, or even double-peaked in the case of counter-rotating disks). 

A very general way of representing any function is via a histogram, as in the \Nukers code \citep{Gebhardt2000}, or as a cubic spline, as in \citet{Merritt1997}; both are special cases of a B-spline basis set. In either case, the number of grid points in the velocity space and associated free parameters (values of the function at these points) is rather large ($10-50$), and a maximum penalized likelihood method is used to recover only the significant features in the data and to prevent overfitting. Consequently, the effective number of free parameters is lower (in the limit of infinite smoothing, only two -- mean and dispersion), and the uncertainties on the function values have a nontrivial correlation matrix, which must be taken into account when fitting a dynamical model. For instance, \citet{Gebhardt2000, Gebhardt2003} estimate that only half of their 13 velocity bins are independent. \citet{Houghton2006} introduce a method for converting the histogrammed representation of a LOSVD into another set of numerically constructed basis functions (which they call ``eigen velocity profiles''), which orthogonalizes the error correlation matrix, but this approach has seen very little practical usage.

The more commonly used alternative is the Gauss--Hermite (GH) expansion \citep{vdMarel1993, Gerhard1993}, in which the LOSVD is given by
\begin{equation}  \label{eq:gh}
g(v) = \frac{\Xi}{\sqrt{2\pi}\;s} \exp\left[-\frac{(v-v_0)^2}{2s^2}\right] \times
\sum_{m=0}^{M}\! h_m\,\mathcal H_m \left(\!\frac{v-v_0}{s}\!\right)\!,
\end{equation}
where $\Xi$ is the overall amplitude, $v_0$ and $s$ are the center and width of the Gaussian function, $\mathcal H_m$ are (modified) Hermite polynomials, and $h_m$ are the coefficients of expansion (it is convenient to normalize $\Xi$ so that $h_0=1$).
If all coefficients with $m>0$ are zero, this corresponds to a pure Gaussian function with mean $v_0$ and dispersion $s$; however, in general both the mean velocity $\overline v$ and the dispersion $\sigma$ depend on all coefficients $h_m$ and may differ from $v_0$ and $s$, respectively.

It is important to keep in mind that $v_0$ and $s$ are parameters of the basis set (in the same way as grid points in velocity space for a histogram representation), while $h_m$ are the coefficients of expansion of a particular function, so they are conceptually different. In other words, a given function $f(v)$ can be approximated by a GH series for any choice of $v_0$ and $s$, but of course the coefficients $h_m$ would be different for each choice, and usually the goal is to build a good approximation with as few terms as possible. It is easy to show that if and only if $v_0$ and $s$ are chosen to be the mean and width of the best-fit Gaussian approximation of the function $f(v)$, then $h_1=h_2=0$. The GH basis set is orthogonal (for a fixed choice of $v_0,s$), meaning that one may construct truncated expansions with different orders $M$, and all coefficients with $m\le M$ will not depend on the choice of $M$. 
On the other hand, when using the GH parametrization of the LOSVD in spectral fitting, the function $f(v)$ to be approximated is unknown a priori, and it is common to vary the parameters $\Xi,v_0,s$ and coefficients $h_3..h_M$ simultaneously to obtain the best fit, while still keeping $h_0=1,h_1=h_2=0$ -- this is the approach used in the popular spectral fitting code \textsc{ppxf} \citep{Cappellari2004}. In this case, the best-fit values $v_0,s$ and all coefficients $h_m$ do depend on the truncation order $M$. 

The advantage of the GH parametrization is that the uncertainties are nearly uncorrelated (at least when the GH coefficients $h_3..h_M$ are small, see Equation~11 in \citealt{vdMarel1993}). However, dealing with uncertainties in $v_0,s$ is awkward, because these are nonlinear parameters of the basis set, rather than coefficients of the linear expansion of the LOSVD. Therefore, it is customary to treat $v_0, s$ as fixed parameters, and translate their uncertainties $\epsilon_{v_0}$, $\epsilon_s$ into the uncertainties on $h_1, h_2$, whose measured values are zero: in the linear approximation, $\epsilon_{h_1} = \epsilon_{v_0} / (\sqrt{2}\,s)$ and $\epsilon_{h_2} = \epsilon_s / (\sqrt{2}\,s)$ \citep[Equation~12 in][]{Rix1997}.

We note that in practice, the normalization of the LOSVD is not known from observations, due to uncertainties regarding sky subtraction and other factors. Hence it must be determined from the surface density profile, convolved with the PSF of the spectroscopic instrument and integrated over each aperture. Even though the self-consistency constraints on the 3D density profile imply that its projection should also follow the observational surface brightness profile, previous studies usually found it beneficial to constrain it separately. We follow this practice, computing the normalization of the LOSVD in each aperture from the surface density profile of the model, convolved with the PSF. It is then required to be reproduced by the weighted sum of orbit contributions to each aperture (orbit LOSVD collapsed along the velocity axis).  

In our code, LOSVDs of each orbit are first recorded as three-dimensional datacubes (two image-plane coordinates and the velocity axis), and represented in terms of a basis set of tensor-product B-splines with a degree ranging from 0 (histograms) to 3 (cubic splines), chosen by the user. These B-splines are defined by grids separately in each dimension; the velocity axis is illustrated in Figure~\ref{fig:bsplines}, central panel. For each point sampled from the trajectory during orbit integration, we accumulate its projection onto each basis function in all three dimensions. Spatial convolution is performed in terms of B-spline representation, and then the LOSVDs are rebinned in the two spatial dimensions onto the set of apertures (defined by arbitrary polygons in the image plane) in which observations were recorded (e.g., elements of a long slit, Voronoi bins, etc.). The convolution and rebinning are expressed as a single matrix-vector multiplication, which is very efficient on modern processors. For each orbit, a two-dimensional array of coefficients of B-spline expansion is stored in the orbit library (one dimension is the velocity axis, the other is the index of the aperture). Technical details of this approach are explained in Appendix~\ref{sec:Bspline}. When fitting the model to kinematic observables, we further convert this B-spline representation, possibly scaled by $\sqrt{\Upsilon}$ as explained in Section~\ref{sec:ML}, into the required form (histogram or GH series, as shown in Figure~\ref{fig:bsplines}, right panel) in each aperture.

In other implementations of the Schwarzschild method, LOSVDs of the orbit library are usually represented by histograms, which are a special case of B-splines (of degree 0). Spatial convolution is performed either by fast Fourier transform (\Nukers and \Monica codes) or by randomly perturbing the coordinates of points stored during orbit integration (\Leiden and \Remco codes). Our approach is significantly more efficient and more accurate, when used with 2nd or 3rd-degree B-splines, see Figure~\ref{fig:bspline_losvd} in Appendix~\ref{sec:accuracy}. On the one hand, higher-degree basis functions enforce greater smoothness of the LOSVD. On the other hand, at a fixed grid spacing, they can represent steeper gradients, which means that one can use coarser grids (with spacing comparable to the spaxel size), and still resolve all relevant features (with accuracy of order 1\%), while saving storage space and computational time.

Regardless of whether the observed LOSVD is represented in terms of a histogram or a GH series, the LOSVD of each $i$-th orbit can be expanded in the same basis set, and the resulting coefficients $u_{in}$ form the matrix of linear equations to be fitted in the least-square sense (Section~\ref{sec:optimization}). In doing so, we may actually use a higher order of GH expansion (e.g., $M=10$) than the observed one (typically 4 or 6), requiring the higher-order GH terms to be zero with some fiducial uncertainty of a few percent. This reduces the propensity of the model to produce unphysically jagged LOSVDs. 

Alternatively, a linear-superposition problem may be formulated for the ``classical'' (as opposed to GH) moments of the LOSVD, namely, the mean velocity $\overline v$, its full second moment $\overline{v^2} \equiv \overline v^2 + \sigma^2$, and possibly higher terms such as $\overline{v^4}$. These quantities are easily calculated from the B-spline representation of each orbit's LOSVD, although in the spherical Schwarzschild code of \citet{Breddels2013} they are computed directly during orbit integration. Because monomials of $v$ form yet another basis set, the solution is still linear in orbit weights (this would not be so, had we used $\sigma$ instead of $\overline{v^2}$ as the observational constraint; see \citealt{Zhao1996}). In practice, GH moments are somewhat better determined observationally, being less sensitive to the often poorly measured wings of the LOSVD than the ``classical'' moments.

\subsection{Solution of the optimization problem}  \label{sec:optimization}

After the orbit library has been constructed, and possibly rescaled in velocity (Section~\ref{sec:ML}), the orbit weights $\boldsymbol w \equiv \{w_i\}_{i=1}^{N_\mathrm{orb}}$ are determined as the best-fit solution to the constrained optimization problem. Namely, we write down the objective function $\mathcal Q(\boldsymbol w)$ to be minimized:
\begin{equation}  \label{eq:objective_function}
\mathcal Q \equiv \sum_{n=1}^{N_\mathrm{obs}} 
\left( \frac{ \sum_{i=1}^{N_\mathrm{orb}} w_i\, u_{in} - U_n }{ \epsilon_{U_n} } \right)^2 +
\mathcal S( \boldsymbol w ),
\end{equation}
where $U_n$ are the values of observational constraints, $\epsilon_{U_n}$ are their measurement uncertainties, $u_{in}$ are the same observables recorded for each orbit, and $\mathcal S(\boldsymbol w)$ is the optional regularization term discussed in Section~\ref{sec:regularization}. The solution must satisfy the non-negativity constraints
\begin{equation}  \label{eq:nonnegativity}
w_i \ge 0,\quad i=1..N_\mathrm{orb},
\end{equation}
and possibly some other equality constraints (e.g., the self-consistency constraints for the intrinsic density profile, as described in Section~\ref{sec:density_constraints}):
\begin{equation}  \label{eq:equality_cons}
\sum\nolimits_{i=1}^{N_\mathrm{orb}} w_i\,m_{ik} = M_k, \quad k=1..N_\mathrm{cons}.
\end{equation}

Without these constraints and ignoring for the moment the regularization term, Equation~\ref{eq:objective_function} is the classical non-negative linear least-square problem (NNLS), which is usually solved using the method of \citet{Lawson1974} -- this approach is followed in \Monica, \Leiden and \Remco codes. However, the venerable old algorithm is far from being optimal in performance, and cannot deal with equality constraints.
Because of the latter reason, many studies treat the intrinsic density constraints as another set of approximate constraints, assigning them some arbitrary but small relative uncertainties (typically $1-2$\%). However, this complicates the interpretation of the confidence intervals on the model parameters, because the $\chi^2$ values have contributions from both observable quantities and the additional intrinsic density constraints.

Alternatively, Equations~\ref{eq:objective_function}--\ref{eq:equality_cons} can be reformulated as a quadratic programming problem \citep[cf.][]{Dejonghe1989}, introducing an auxiliary vector of $N_\mathrm{obs}$ slack variables $s_n$ and associated equality constraints
\begin{equation}  \label{eq:slackness_cons}
\sum\nolimits_{i=1}^{N_\mathrm{orb}} w_i\, u_{in} + s_n = U_n, \quad n=1..N_\mathrm{obs} .
\end{equation}

Combined with Equation~\ref{eq:equality_cons}, we have a system of $N_\mathrm{cons}+N_\mathrm{obs}$ linear equations for $N_\mathrm{orb}+N_\mathrm{obs}$ variables, satisfying the non-negativity constraints (\ref{eq:nonnegativity}), and the objective function becomes
\begin{equation}
\mathcal Q = 
\sum_{n=1}^{N_\mathrm{obs}} \left(\frac{s_n}{\epsilon_{U_n}}\right)^2 + \mathcal S(\boldsymbol w)
\equiv \chi^2 + \mathcal S(\boldsymbol w).
\end{equation}

After experimenting with many black-box quadratic programming solvers, we found the open-source \textsc{cvxopt}\footnote{\url{http://cvxopt.org}} library to be most efficient for our purposes. Most commercial solvers such as \textsc{cplex}, \textsc{gurobi}, \textsc{mosek} and \textsc{galahad} (the latter was used for some time in the \Remco code) are optimized to deal with sparse matrices of linear and quadratic constraints. In our case, the matrix of linear constraints is typically quite dense, but the matrix of quadratic constraints is diagonal (if the regularization term $\mathcal S$ is just a sum of squared orbit weights). We have modified the \textsc{cvxopt} library to take advantage of this structure of the problem, and it can utilize highly optimized state-of-the-art dense linear algebra libraries such as \textsc{OpenBLAS}, taking advantage of both multi-core parallelization and the SIMD instruction set of modern CPUs. For instance, with 64 threads it reaches a peak performance of $\sim 100$~Gflops per CPU core on a 2~GHz Intel Xeon processor, i.e. 50 flops per CPU cycle -- something that would be nearly impossible to achieve in programs written in a conventional coding style, without extensive architecture-specific low-level fragments.

The massive speedup of the optimization procedure is one of the key improvements in our code, which allows it to solve a problem with $\mathcal O(10^5)$ orbits and $\mathcal O(10^4)$ constraints in just a few minutes on a high-end multicore CPU.

We note, however, that this efficient optimization solver can be used only when the quadratic objective function is diagonal, or in other words, when there are no correlations between observational errors. Moreover, when the linear-superposition method is used to fit discrete-kinematical data (velocities of individual stars rather than LOSVDs), as in \citet{Chaname2008, Magorrian2019}, the objective function is not quadratic in orbit weights anymore, and the problem requires a general nonlinear optimization solver.

\subsection{Regularization}  \label{sec:regularization}

Since the number of orbits in Schwarzschild models is usually much larger than the number of observational constraints, the solution for the orbit weights is highly non-unique. \citet{Magorrian2006} argues that if one is interested in comparing the likelihood of different potentials, not the DF itself (essentially the orbit weights), then one needs to marginalize over all possible DFs which are allowed by each potential. Naturally, this is almost infeasible computationally, although he demonstrates the possibility of performing such a marginalization in a toy model. More recently, \citet{Bovy2018} revisited this approach in the context of the made-to-measure method applied to a toy harmonic potential. Still, a full Bayesian treatment of the DF as a set of nuisance parameters in a realistic potential and with many thousand orbits is a remote goal at the moment.

Putting aside the question of marginalization, one has to be content with a single best-fit solution for orbit weights for the given choice of potential parameters. The dynamical inverse problem -- determination of the DF from its noisy projection into the observable space -- is a classic example of an ill-conditioned problem, which usually requires some sort of regularization technique to produce a meaningful solution (see the discussion in, e.g., \citealt{Merritt1993}).

There are two commonly used approaches to regularization in the context of Schwarzschild models: local and global. In the first case, one seeks a solution in which nearby orbits in the space of integrals of motion would preferrably have similar weights. This is usually implemented as a penalty term $\mathcal S(\boldsymbol w)$ in the objective function (\ref{eq:objective_function}) that is  proportional to the squared second derivative of the DF as a function of integrals of motion. In absence of a complete set of classical integrals (essentially in any non-spherical system), they are substituted by the initial positions of orbits in a regularly-structured start space (Section~\ref{sec:orbit_library}). This approach is followed in the \Leiden, \Remco and \Monica codes. 
However, one disadvantage of our random sampling scheme for assigning initial conditions is that it does not provide any measure of proximity of orbits in the integral space. In any case, in the local approach the penalty function $\mathcal S$ is a bilinear form of the solution vector with a non-diagonal matrix, which would prevent the possibility of using the optimized quadratic-programming solver.

The second approach dispenses with the requirement that orbit weights should be similar locally, and instead imposes integral constraints on the overall distribution of orbit weights. \citet{Richstone1988} introduced the maximum-entropy approach in a general context, which was subsequently adopted in the \Nukers code. The regularization penalty term in the objective function is proportional to the Boltzmann entropy $-\int f(\Bx,\Bv)\, \ln f(\Bx,\Bv)\; \d^3\Bx\,d^3\Bv$, or expressed in terms of orbit weights, $\sum_{i=1}^{N_\mathrm{orb}} (w_i / \tilde w_i) \, \ln(w_i / \tilde w_i)$, where $\tilde w_i$ is the prior on the orbit weight. In the \Nukers code, $\tilde w_i$ are the phase volumes associated with each orbit, computed from the Voronoi tesselation of the surface of section \citep{Thomas2004}.
Increasing this penalty term makes the distribution of orbit weights more uniform.  
The non-linear functional form of this penalty means that a quadratic programming method is not applicable; instead, the solution is obtained by a more general Newton's method with special adaptations to enforce non-negativity of the solution vector. 
On the other hand, Boltzmann entropy doesn't play a special role in this context, and any convex function would produce a similar regularizing effect. Accordingly, we choose to use a diagonal quadratic penalty (similar to that used by \citealt{Merritt1996}):
\begin{equation}  \label{eq:regul}
S = \lambda\, N_\mathrm{orb}^{-1}\, \sum\nolimits_{i=1}^{N_\mathrm{orb}} (w_i / \tilde w_i)^2,
\end{equation}
where again $\tilde w_i$ are priors on orbit weights -- in our random sampling approach for the generation of initial conditions, these values are all identical and equal to $M / N_\mathrm{orb}$, but a more sophisticated choice of priors is also possible.

In all regularization schemes, a free parameter (called $\lambda$ in the above equation) determines the relative importance of regularization penalty term in the overall objective function. The standard practice is to choose it in such a way that the quality of fit does not significantly deteriorate, or in other words, $\chi^2$ (the first term in Equation~\ref{eq:objective_function}) increases by $\mathcal O(1)$ compared to the case without regularization. Alternatively, one may determine the optimal value of $\lambda$ by cross-validation (McDermid et al., unpublished). We find that values of $\lambda \sim \mathcal O(1)$ produce adequate results, but defer a more thorough exploration of the regularization to a future study.

\subsection{Analysis of the orbital structure}

We may explore the internal structure of the best-fit model in several ways. 
For each orbit, we compute the intrinsic kinematic properties such as the velocity moments discretized on a suitable 3D grid. These quantities are then multiplied by the orbit weights in the solution and summed up to obtain the overall profiles. 

For spheroidal systems, it is instructive to consider the velocity anisotropy coefficient $\beta \equiv 1 - (\sigma_\theta^2 + \sigma_\phi^2) / (2 \sigma_r^2)$. 
Figure~1 in \citet{IAU} illustrates a well-known fact that models with the same observed kinematics but different potentials have rather different profiles of $\beta(r)$, generalizing the anisotropy inversion approach \citep{Binney1982} to non-spherical systems. It also demonstrates that the intrinsic kinematic properties may change rather drastically outside the radius where the model is constrained by kinematic observations. Since the photometry is usually available out to larger distances than kinematics, the intrinsic density profiles behave more regularly.

Another way of looking into the orbit distribution is provided by analyzing the weights of orbits as functions of integrals of motion (or their approximations) such as $E$, time-averaged inclination of the orbital plane $\cos i \equiv \overline{L_z/L}$, or the orbit circularity parameter $\overline{L_z}/L_\mathrm{circ}(E)$ introduced in \citet{Zhu2018}. 
In Section~\ref{sec:tests_barred}, we demonstrate that the Schwarzschild models are able to recover the orbit distribution for the best-fit (correct) values of parameters, and concur with the authors of the \Leiden and \Remco codes that it is advisable to use a nonzero regularization coefficient $\lambda$ to reduce fluctuations and unrealistic sudden changes in the orbital structure.

A more sophisticated analysis of the orbital structure of triaxial systems is possible with tools such as frequency maps \citep{Valluri1998,Valluri2016}, which highlight various resonant families. These tools a were part of the earlier version of our Schwarzschild code (\textsc{Smile}; \citealt{Vasiliev2013}) but are not yet included in the current version. This kind of analysis is especially interesting in application to bars (see e.g.\ a similar study of \citealt{Portail2015} in the context of M2M models). 

Finally, an orbit-superposition model may be converted into an $N$-body model, by sampling a number of points from each orbit in proportion to its weight. This could be useful, e.g., for testing the stability of a given solution. Of course, the Schwarzschild method itself can serve as a way of creating equilibrium models with prescribed density profiles, not necessarily constrained by any observations; \textsc{Smile} has been used in this context \citep{Vasiliev2012,Vasiliev2015}.

\subsection{Implementation and workflow}

The present Schwarzschild modelling code forms part of the \Agama framework for galaxy modelling, together with other methods based on DFs in action space, Jeans equations, etc. It shares many aspects (such as the collection of potential models or the representation of velocity distribution in terms of B-splines) with those methods, but many tasks are performed somewhat differently. For instance, in DF-based methods, any observable quantity such as an LOSVD is computed directly from DF for any point on the sky, whereas in the orbit-superposition method one needs to specify the sky-plane apertures before building the orbit library and obtaining the solution of the optimization problem. Similarly, sampling $N$-body particles from a DF can be done at any time, but in the Schwarzschild method these samples must be collected during orbit integration.

The computational core of the \Agama library is written in C++, but the top-level workflow of the Schwarzschild modelling method is implemented in Python for a greater flexibility. The computationally intensive functions -- construction of potential and density models, preparation of initial conditions, definition of density and LOSVD targets, orbit integration, computation of GH moments, quadratic optimization solver, conversion of the orbit library into an $N$-body snapshot --  are contained in the core of the library and are accessible through its Python interface. These operations are all OpenMP-parallelized, hence can use all available CPU cores of a single machine.
The choice of density/potential components, various parameters of the model, and data acquisition and preparation are usually specific to each galaxy, so should be provided by the user in the Python script. 

The workflow usually consists in defining a model (all parameters of potential, geometry, etc. up to an overall mass-to-light ratio $\Upsilon$), reading and preparing observational constraints, constructing the orbit library, and solving the optimization problem for different choices of $\Upsilon$ using the same orbits but rescaling the velocity, as explained in Section~\ref{sec:ML}.
Typically $\Upsilon$ is not the only free parameter in the model, and separate orbit libraries should be constructed for different choices of all other parameters; these independent subsets of models can be run in parallel on different machines. The values of $\chi^2$ for all models are then plotted in a common parameter space, and if necessary, marginalized over some dimensions (e.g., M/L) to obtain final confidence intervals for parameters of interest (e.g., SMBH mass).

A more detailed description of the code is included in the \Agama reference documentation \citep{Vasiliev2018}. We provide examples of the entire workflow, and an interactive Python script for analyzing the modelling results (plotting the contours of $\chi^2$ in the parameter space, maps of $v_0$, $s$ and higher GH moments for different models and the original data, examining LOSVDs in individual apertures, etc.)

\section{Tests}  \label{sec:tests}

\subsection{Generation of mock datasets}

In order to validate the code, we prepare mock input data with parameters similar to the commonly used observational datasets.

We use several different DF-based or $N$-body models of disk galaxies with and without bars, which will be described in more detail in subsequent sections. We choose dimensional scaling units in such a way as to mimic a Milky Way-sized galaxy, with stellar mass $\sim 5\times10^{10}\,M_\odot$, half-light radius of $\sim 3$~kpc, and peak circular velocity of $\sim 200-250$~km/s. We place the galaxy at a fiducial distance 20~Mpc, hence $1''\simeq 100$~pc.

Each $N$-body model is used to create several mock datasets with different inclinations and, in the case of triaxial models, orientations of the major axis of the bar. 
As explained in Section~\ref{sec:deprojection}, inferring the 3D shape of the galaxy from the projected surface brightness profile is a difficult and underconstrained problem, although adding the kinematic information may lift some degeneracies, as explored by \citet{vdBosch2009} in the context of triaxial spheroidal galaxies. We leave this topic for a future study, and in the present paper we assume that the 3D shape of the galaxy is known, thereby side-stepping the deprojection problem. In practice, we construct a smooth non-parametric representation of the 3D potential on a cylindrical grid, as explained in Section~\ref{sec:potential}, directly from the input $N$-body snapshot, and only vary the overall normalization $\Upsilon$ during the fit, as explained in Section~\ref{sec:ML}.

We construct two kinematic datasets: low-resolution (LR) dataset covering a large spatial region, roughly up to one half-light radius, and high-resolution (HR) dataset covering only the central region, but with a much smaller PSF. For the former, we adopt parameters similar to those of the large-scale IFU instruments such as SAURON/ATLAS3D 
or VIMOS 
(see also Table~1 in \citealt{Zou2019} for a compilation of properties of various IFU instruments): field-of-view (FoV) $60''\times60''$, pixel size $1.0''$, spatial resolution (width of the Gaussian PSF) $1''$. For the HR dataset, we take the typical parameters of AO-assisted IFU such as NIFS or SINFONI: FoV $2''\times2''$, pixel size $0.05''$, PSF width $0.1''$. In both cases the IFU is centered on the galaxy, and the kinematic data are point-symmetrized so that the observed LOSVD $\mathcal F(X,Y,V) = \mathcal F(-X,-Y,-V)$. This allows one to use only half of the image plane, irrespective of the orientation of the IFU, even for barred galaxies. We use the Voronoi binning approach \citep{Cappellari2003} to group the pixels into $\sim50-100$ apertures in each dataset, roughly maintaining a constant total flux per bin.

The LOSVDs in each bin are computed either from $N$-body particles or directly from the analytic DFs of the models, using the same sequence of operations as when sampling points from orbits during the Schwarzschild modelling. We then convert the LOSVDs into the GH representation with 6 GH moments. The intrinsic discreteness (Poisson) noise is fairly low when using high-resolution $N$-body simulations, or negligible when using analytic DFs. 
We assign the formal uncertainties typical of the modern instruments: $\epsilon_{v_0}, \epsilon_s = 5$~km/s, and $\epsilon_{h_3\dots h_6} = 0.02$. We use both the ``clean'' mock kinematic maps, with negligibly low Poisson noise, and ``noisy'' maps, in which each quantity is perturbed by a Gaussian error with the quoted standard deviation.

\subsection{Axisymmetric disk models}  \label{sec:tests_axisym}

\begin{figure*}[t]
\includegraphics{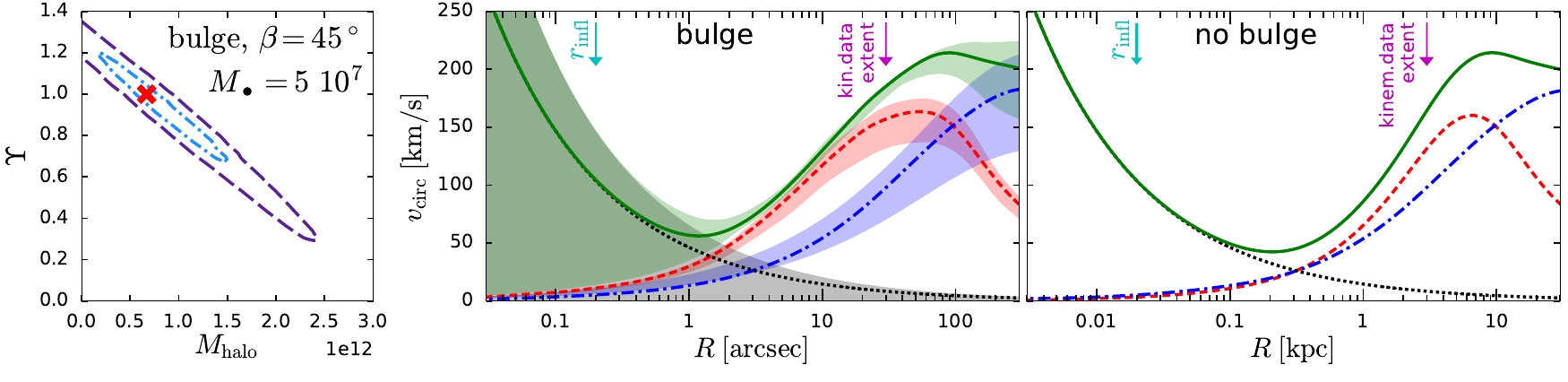}
\includegraphics{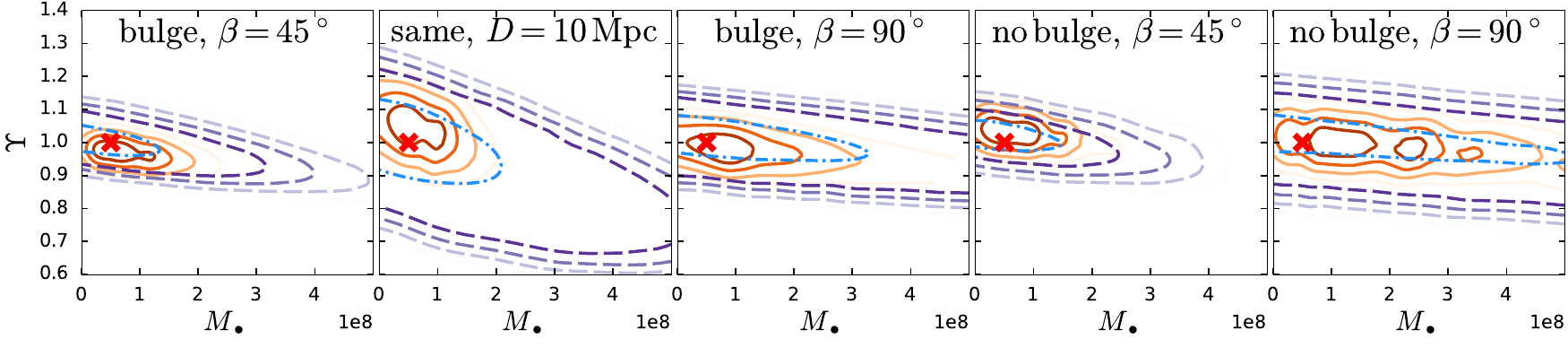}
\includegraphics{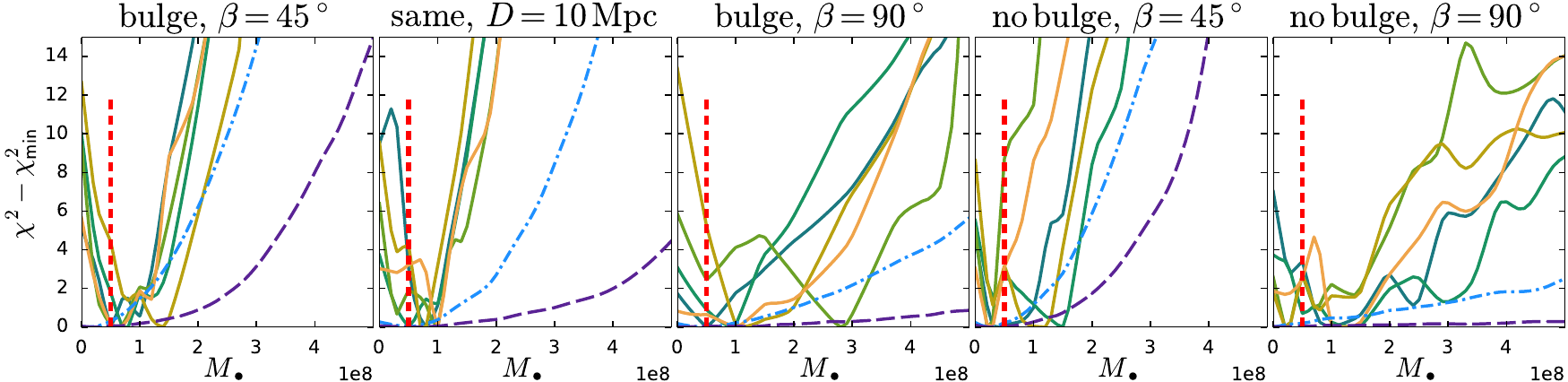}
\caption{
Properties of axisymmetric mock datasets and modelling results.\protect\\
\textit{Top row:} circular velocity curves of the model with a central bulge (center panel) and without a bulge (right panel). The contributions of disk (+bulge), halo, central SMBH, and the total circular velocity $v_\mathrm{circ}\equiv \sqrt{R\, \d\Phi/\d R}$ are shown by red dashed, blue dot-dashed, black dotted, and green solid lines, respectively, as functions of the distance from the galaxy center (1''=0.1~kpc).
The left panel shows the degeneracy between stellar M/L ratio $\Upsilon$ and the mass of the dark halo (for a fixed $M_\bullet$ equal to the true value), with the true values marked by a cross. All models within the ellipses have $\chi^2<2.3$ (when using noiseless data, the absolute values of $\chi^2$ don't have any special meaning, but these models are essentially perfect fits to the data). The outer dashed ellipse shows unregularized models, and the inner dot-dashed one -- models with relatively large regularization coefficient $\lambda=10$. The circular velocity curves of the latter series of models are shown as shaded regions in the middle panel: both disk and halo contributions have much larger uncertainty than the total circular velocity within the range of radii probed by the data (shown as a magenta vertical arrow). The SMBH radius of influence is marked by a cyan vertical arrow; note that it is significantly smaller than the radius at which the gravity is dominated by the SMBH.
\protect\\
\textit{Middle row:} contours of $\Delta\chi^2$ (difference in $\chi^2$ between the given model and the best-fit model) in the parameter space ($M_\bullet$ vs. $\Upsilon$), for several choices of models: with or without a bulge, inclination angle $\beta=90^\circ$ (edge-on) or $45^\circ$, and one model placed at a twice closer distance (10~Mpc vs.\ the default 20~Mpc).
Purple dashed lines: noiseless data and unregularized models ($\Delta\chi^2=2.3, 6.2, 11.8$); 
blue dot-dashed: same data, $\lambda=10$ (only the inner contour shown); orange solid: one realization of noise, $\lambda=10$. The true parameters are shown by a red cross.
\protect\\
\textit{Bottom row:} contours of $\Delta\chi^2$ as functions of $M_\bullet$, marginalized over the other parameter ($\Upsilon$) for the same models as in the previous row. Wider curves are for the noiseless models (magenta dashed: without regularization, blue dot-dashed: with the regularization coefficient $\lambda=10$), while the other colored curves show five different realizations of noise in each case (all with $\lambda=10$, although the curves look rather similar for other choices of $\lambda$). Vertical red dashed line marks the true value of $M_\bullet$.
}  \label{fig:axi_models}
\end{figure*}

We first test the new Schwarzschild code in the axisymmetric regime. 
For this exercise, we construct models defined by distribution functions in action space, using the iterative approach implemented in \Agama. The models have a nearly exponential disk with scale length 3~kpc, scale height 0.3~kpc, optionally a central S\'ersic bulge with scale radius $\sim 1$~kpc, a nearly-NFW halo, and a central BH. The total mass of the disk and the bulge is $5\times10^{10}\,M_\odot$ (the bulge, if present, contains 20\% of this mass), the contribution of the halo to the rotation curve reaches 50\% at $R\simeq 10$~kpc, and the central BH has a mass $10^{-3}\times M_\mathrm{disk+bulge}$. Figure~\ref{fig:axi_models}, top row, shows the rotation curves of the models, which are similar to that of the Milky Way.
The central velocity dispersion is $\sigma \sim100$~km/s, corresponding to the radius of influence $r_\mathrm{infl} \equiv G\,M_\bullet / \sigma^2 \simeq 20\mbox{ pc} = 0.2''$, twice larger than the HR PSF width $0.1''$. These values are typical for recent studies of SMBH in galaxies of similar $\sigma$, distance and $M_\bullet$ (e.g., \citealt{Krajnovic2018}, or Tables~1 and 5 in \citealt{Thater2019}).

Since axisymmetric models have fourfold symmetry when the kinematic datacube is aligned with the major axis, we use only one quadrant on the sky plane. Both LR and HR datasets contain 50 Voronoi bins, i.e. 300 kinematic constraints. We also fit the intrinsic 3D density profile discretized on a cylindrical grid with 300 constraints, and the surface density profile (LOSVD integrated along the velocity dimension in each aperture), requiring an exact fit in both cases. Hence the $\chi^2$ values reflect only the difference in kinematic constraints. We use 20\,000 orbits for all models ($20-30\times$ higher than the number of total or kinematic constraints), and vary the regularization parameter $\lambda$ in (\ref{eq:regul}) between 0 (no regularization) and 10 (relatively strong one).

First, we run the code on the ``clean'' (noise-free) mock data, while still using the formally assigned error bars. Of course, in this case the values of $\chi^2$ do not have any statistical meaning -- only the region of the parameter space with essentially perfect fits ($\chi^2\approx 0$) is significant, as it illustrates the intrinsic flexibility and degeneracy of the models.

The total potential is composed of the stellar disk (including the bulge), the dark halo, and the central BH. For the halo, we use a spherical NFW profile with a fixed scale radius of 20~kpc and adjustable normalization, even though it is somewhat different from the actual halo density profile of our models.
We find that with the adopted spatial coverage of the kinematic maps (up to 1 half-light radius), we are not able to disentangle the contribution of the disk and the halo to the total potential: the halo normalization is strongly degenerate with the stellar $M/L$ ratio $\Upsilon$ (Figure~\ref{fig:axi_models}, top left panel). This remains true even for the datasets with added noise, so we conclude that a larger FoV would be needed to constrain the halo properties. We fix the halo normalization to the true value henceforth. 

We then focus on the two remaining parameters -- $\Upsilon$ and $M_\bullet$. Figure~\ref{fig:axi_models}, middle row, shows that in the noise-free case, the constraints on $M_\bullet$ are very weak: any value between 0 and $5-10\times$ the true BH mass is equally consistent with the data. Such flat-bottomed $\chi^2$ contours have been previously demonstrated by \citet{Valluri2004} in a similar context. This is not unexpected, since the models are very flexible, and can accommodate a wide range of the BH mass by counterbalancing changes in the orbital structure at larger radii.

A closer examination reveals that the models at the edge of the allowed parameter space are less realistic, having large disparity in orbit weights and significant fluctuations in the kinematic structure outside the range of radii constrained by the data. The first two panels in the middle row of Figure~\ref{fig:axi_models}, or Figure~1 in \citet{IAU}, show that by increasing the regularization parameter $\lambda$ or the spatial coverage of the kinematic maps, one obtains tighter constraints on both $\Upsilon$ and $M_\bullet$ by eliminating extra freedom in orbital structure (see also \citealt{Cretton2004} for a discussion).

In the case of spherical models, \citet{Dejonghe1992} have shown that the 2d DF $f(E,L)$ can be uniquely recovered from the observed LOSVD $\mathcal F(R,v)$ \textit{in the given potential} $\Phi$. They further conjectured that the constraints on $\Phi$ coming from the non-negativity of the recovered DF are quite tight, but did not rigorously demonstrate this. Our experiments suggest that by increasing the spatial coverage, the constraints on $\Phi$ indeed get tighter, possibly even shrinking to a single point in the $\Upsilon-M_\bullet$ plane as the maps cover the entire galaxy. However, this applies only to a restricted two-parameter family of models, and it is not clear if this statement is true or can be proven in a general case. We leave a more thorough exploration of this question for a future study.

\citet{Magorrian2006} confirmed that flat-bottomed $\chi^2$ contours appear in the noise-free case, but argued that with a realistic level of noise, even the intrinsically flexible Schwarzschild models cannot fit the data perfectly, and $\chi^2$ has a well-defined nonzero minimum as a function of model parameters. Figure~\ref{fig:axi_models}, bottom row, illustrates this behaviour for several noise realizations, plotting $\chi^2(M_\bullet)$, marginalized over $\Upsilon$. The curves usually have well-defined minima, but are sometimes quite noisy with multiple local minima. The minimum values of $\chi^2\sim400$ are significantly smaller than the number of constraints ($N_\mathrm{obs}=600$), indicating that the models are still overfitting the noise, regardless of the regularization parameter $\lambda$ (within the range considered). 

The confidence intervals on the model parameters are quoted at a particular level of $\Delta\chi^2 \equiv \chi^2(M_\bullet) - \chi^2_\mathrm{min}$. The standard approach is to use $\Delta\chi^2=1$ as the 68\% (``1$\sigma$'') confidence interval for one degree of freedom ($M_\bullet$ only, after marginalization over the remaining parameters). We see that the true value of $M_\bullet$ is often outside the formal 1$\sigma$ intervals, although still within 2--3 $\sigma$ ($\Delta\chi^2=4$ or 9, correspondingly). Some authors (e.g., \citealt{vdBosch2008} and subsequent papers) argue that the statistical uncertainty in the value of $\chi^2$ itself is $\delta\chi^2 = \sqrt{2\,N_\mathrm{obs}} \gg 1$, and use the latter value to define the confidence intervals. Indeed, the scatter in $\chi^2_\mathrm{min}$ between different noise realizations is consistent with the above estimate; however, for a given noise realization, this scatter is irrelevant for the purpose of determining the confidence intervals. On the other hand, it is universally acknowledged that using $\Delta\chi^2=1$ produces unrealistically small uncertainties. More importantly, the use of a fixed cutoff value of $\Delta\chi^2=1$ for one degree of freedom ignores the fact that the orbit-superposition models have $N_\mathrm{orb}$ hidden free parameters, for which we take only the best-fit values but do not marginalize over them (the point raised by \citealt{Magorrian2006}). It is clear that a more rigorous statistical analysis is needed to robustly determine the confidence intervals on the model parameters and to explore the role and the optimal level of regularization; we leave it for a future study.

Interestingly, the allowed intervals of $\Upsilon$ and $M_\bullet$ become broader when we place the mock galaxy at half the distance of our fiducial models (10~Mpc) while keeping all other parameters unchanged (Figure~\ref{fig:axi_models}, second column in the last two rows). Despite the sphere of influence now being $2\times{}$ larger on the sky plane, $M_\bullet$ is even less well constrained due to a greater freedom available to the model to rearrange the orbits in the outer parts, not covered by the LR dataset. This underlines the need to use the kinematic data across the entire galaxy, even when interested only in the central part of it, or else to put some physically motivated priors on the distribution of orbits not explicitly constrained by observations. 

\subsection{Barred disk models}  \label{sec:tests_barred}

\begin{figure*}[t]
\includegraphics{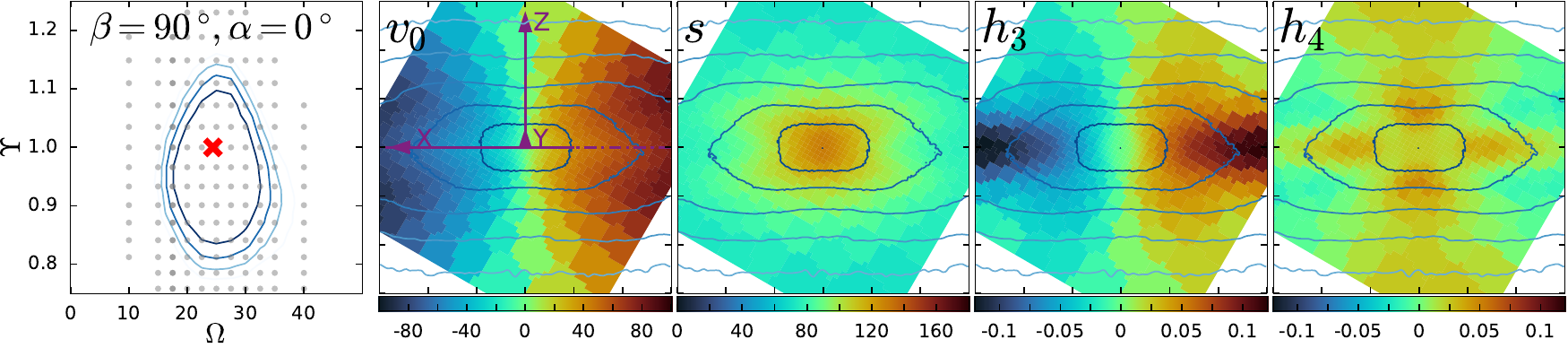}
\includegraphics{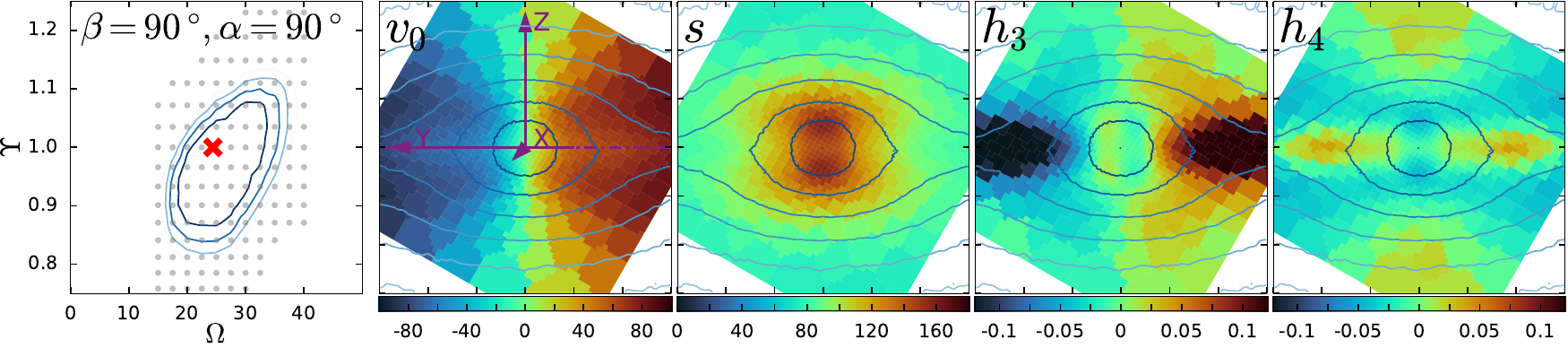}
\includegraphics{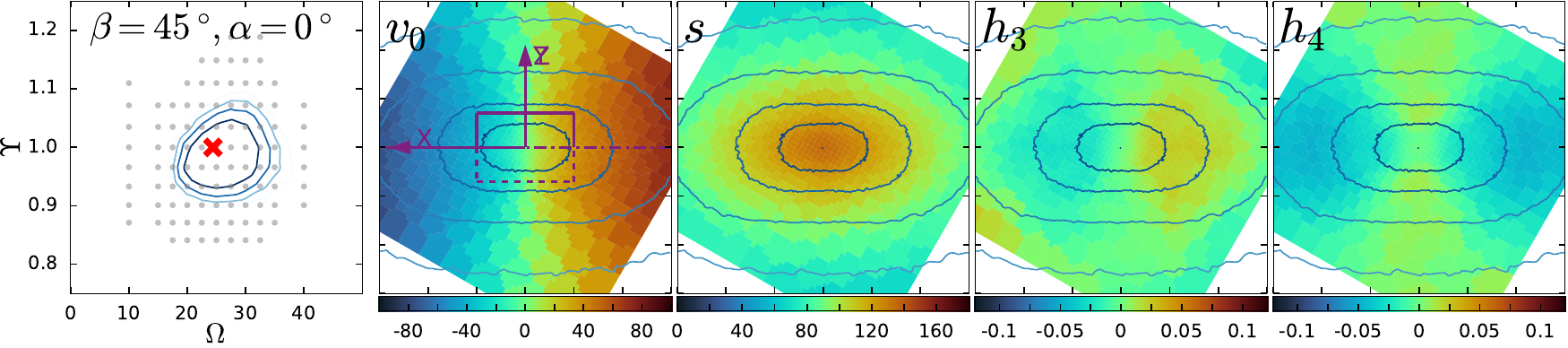}
\includegraphics{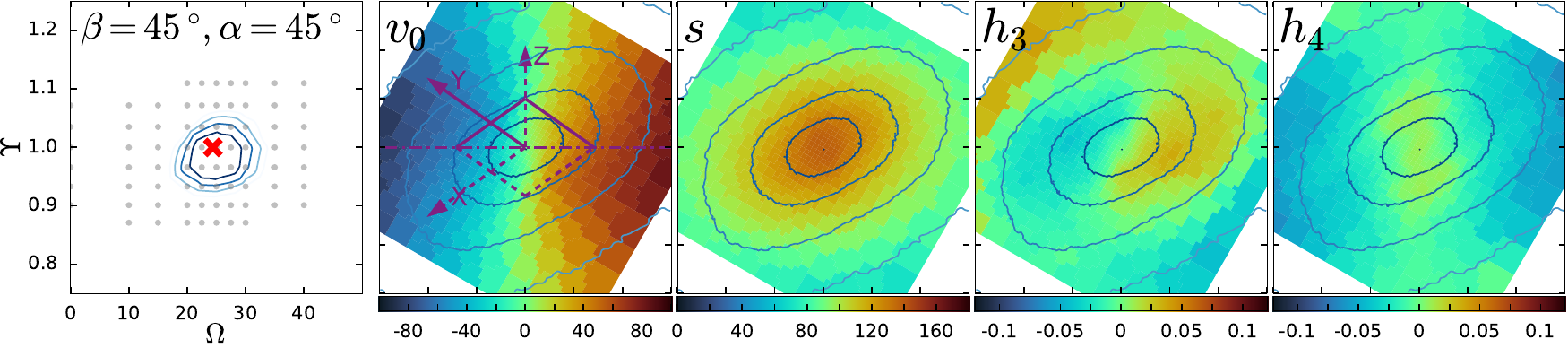}
\includegraphics{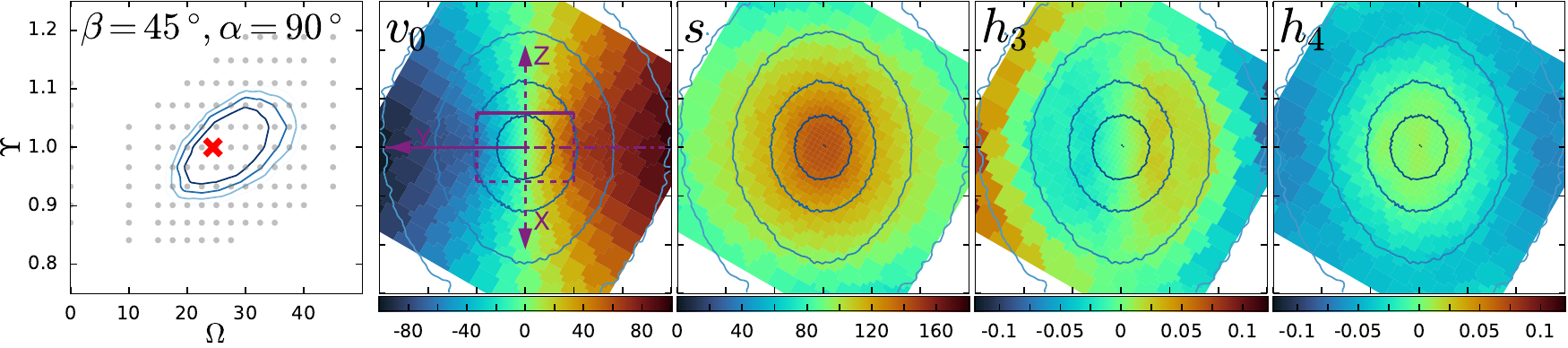}
\vspace*{-2mm}
\caption{Kinematic maps and modelling results for a barred disk galaxy observed at different orientations, parametrized by Euler angles: inclination angle $\beta$ and rotation angle of the bar w.r.t.\ the line of nodes (the latter shown by a horizontal dash-dotted line) $\alpha$, while the angle $\gamma$ is fixed to $30^\circ$. 2nd column shows the intrinsic axes of the system (dashed when behind the image plane) and its equatorial plane by a rectangle. We use 6 GH moments in the models, but show only the first four noise-free maps here, since the features in $h_5,h_6$ are very similar to those in $h_3,h_4$ with inverted sign.\protect\\
Left column shows the ranges of the two model parameters (pattern speed $\Omega$ and mass-to-light ratio $\Upsilon$) consistent with the noiseless data ($\chi^2=2.3,\, 6.2,\, 11.8$; since there is no noise, the minimum value of $\chi^2$ is essentially zero). The true parameters are marked by red cross, and the explored models -- by grey dots. 
\textit{Continued on the next page}
}  \label{fig:barred_models}
\end{figure*}

\begin{figure*}[t]
\figurenum{\arabic{figure} \textit{(continued)}}
\includegraphics{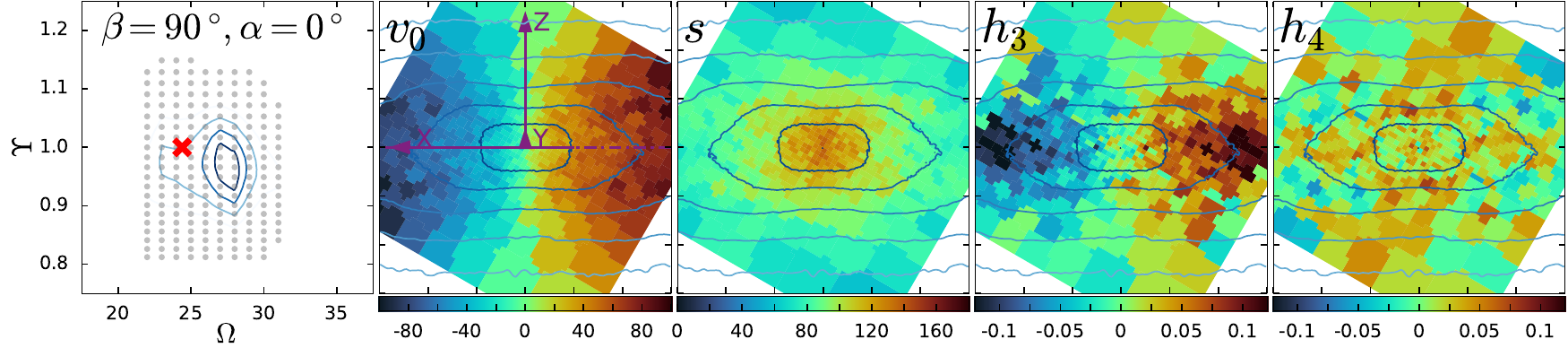}\\[2mm]
\includegraphics{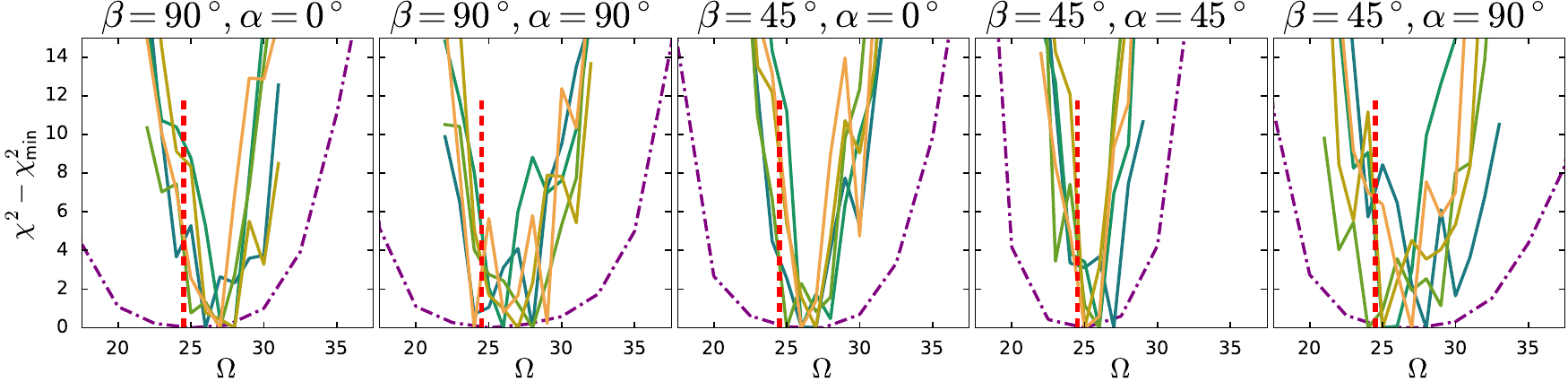}
\caption{\textit{Top row:} same model as in the first row on the previous page, but with added noise. Note a narrower range of $\Omega$ in the left panel. Contours on the maps show the projected density, with contour lines spaced by 1 magnitude (also on the previous page).\protect\\
\textit{Bottom row:} contours of $\Delta\chi^2$ (difference in $\chi^2$ between the given model and the best-fit model) as functions of $\Omega$, marginalized over the other parameter ($\Upsilon$), for the five orientations shown on the previous figure. Wider dot-dashed curves are for the noiseless models, while the other colored curves show five different realizations of noise in each case. Vertical red dashed line marks the true value of $\Omega$.
}  \label{fig:barred_models_noise}
\end{figure*}

\begin{figure*}[t]
\includegraphics{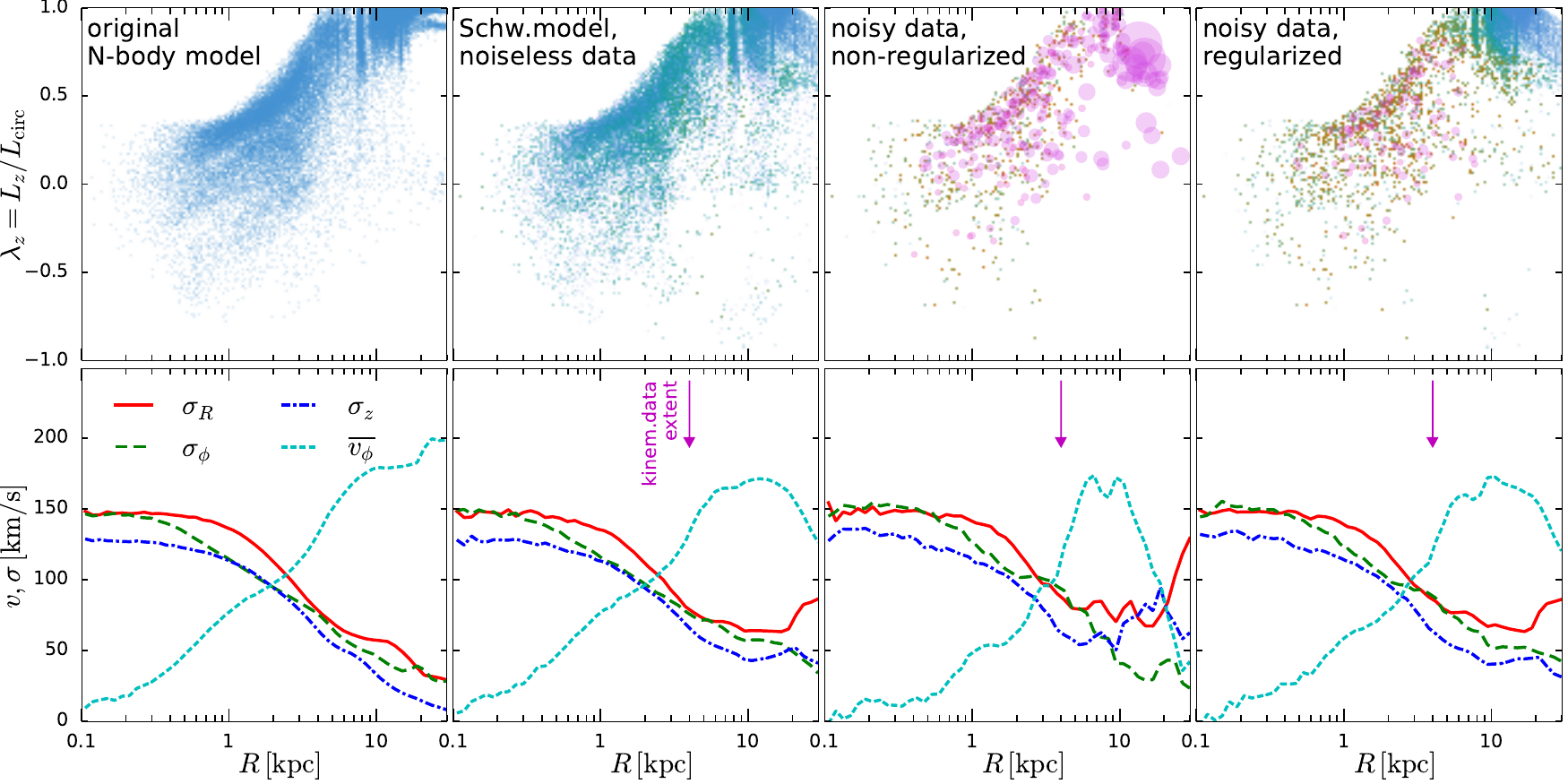}
\caption{\textit{Top row:} orbital structure of the barred models, visualized as the density of orbits in the phase space: mean radius (horizontal axis) vs.\ orbit circularity $\lambda_z \equiv \overline{L_z}/L_\mathrm{circ}(E)$. Left column is the original $N$-body model, and the remaining ones are Schwarzschild model with the correct values of $\Upsilon$ and $\Omega$, and orientation angles $\alpha=45^\circ$, $\beta=45^\circ$ (fourth row on Figure~\ref{fig:barred_models}): noiseless, noisy without regularization, and noisy regularized. Color and size of the points show the orbit weight (blue and green -- low values close to $1/N_\mathrm{orbits}$, red and purple -- large weights). In non-regularized models, there are very few orbits with large weights, while regularized ones have a more uniform weight distribution. Overall, the Schwarzschild models recover the orbital population of the original $N$-body model quite well.\protect\\
\textit{Bottom row:} internal kinematics of the same set of models (left is the original $N$-body model, and the remaining columns are Schwarzschild models). Shown are radial profiles of the mean rotational velocity $\overline v_\phi$ and three components of velocity dispersion tensor $\sigma_{R,\phi,z}$ (averaged over azimuthal angle and vertical direction). Schwarzschild models recover well the internal kinematics within the range of radii constrained by the data (up to $3-5$~kpc, indicated by a purple vertical arrow), although the non-regularized model exhibits large fluctuations.
}  \label{fig:barred_models_structure}
\end{figure*}

We now apply the Schwarzschild code to a barred disk galaxy, using a snapshot from the $N$-body simulation of \citet{Fragkoudi2017} taken after several Gyr of evolution, when an X-shaped bar has fully developed. The $N$-body system is scaled to resemble the Milky Way, has $10^7$ particles in the disk component and is embedded in a live dark matter halo. 
We again place it at a fiducial distance 20~Mpc, and use only the LR datacube with size $60\times60''$, since the models do not contain any central SMBH. In this case, we use one half of the sky plane, because general triaxial models are only point-symmetric, and cover it with 200 Voronoi bins (i.e., 1200 kinematic constraints). We also use twice higher number of intrinsic density constraints (600), with the same cylindrical grid but two angular harmonic terms ($m=0$ and $m=2$), and another 200 surface density constraints. Accordingly, we increase the number of orbits to 50\,000, and use a mild amount of regularization ($\lambda=1$), which is expected not to bias the solution \citep[cf.][]{Valluri2004}.

We consider two inclination angles ($\beta=45^\circ$ and $90^\circ$) and three choices of bar orientation w.r.t.\ the line of nodes (intersection of the galaxy disk and sky planes): $\alpha=0^\circ$ (bar along the projected major axis of the disk), $90^\circ$ (bar perpendicular to the major axis, or seen end-on in case of $90^\circ$ inclination), and $45^\circ$ (intermediate case, when both the photometry and kinematics have twists). We additionally rotate the FoV by $\gamma=30^\circ$ w.r.t.\ the line of nodes in the sky plane. As explained before, we use the true 3D shape of the $N$-body system and only consider one choice of viewing angles (the correct one) for each mock dataset.

Figure~\ref{fig:barred_models} shows the noise-free kinematic maps and the range of parameters $\Upsilon,\Omega$ for which the Schwarzschild models produce essentially perfect fits. As in the previous case, $\Upsilon$ is nearly degenerate with parameters of the dark matter halo profile, so we fix the latter to the initial profile and mass (even though it has likely changed in the course of evolution), hence the correct value of $\Upsilon$ is recovered to within $\lesssim 10\%$, and $\Omega$ is also fairly well constrained. 
Figure~\ref{fig:barred_models_noise} confirms that even in the presence of noise, the best-fit values of $\Upsilon$ and $\Omega$ are close to the true ones ($\Omega$ is systematically overestimated by $\sim10$\%), although the formal uncertainty intervals are likely too tight to be realistic.

The good accuracy of measurement of the pattern speed by the Schwarzschild method is quite remarkable.
The well-known alternative approach for determining $\Omega$ due to \citet{Tremaine1984} deals with a much reduced subset of data: one-dimensional profiles of $\Sigma(l)$ and $v_\mathrm{los}(l)$, measured along the bar, with the coordinate $l$ formally integrated from $-\infty$ to $+\infty$. Due to cancellation of positive and negative contributions to the integrals, this method cannot be applied in symmetric cases (4 out of 5 shown in Figure~\ref{fig:barred_models}): in the edge-on orientation ($\beta=90^\circ$) or when the bar is aligned with the photometric major or minor axes ($\alpha=0^\circ$ or $\alpha=90^\circ$). It is also sensitive to the misalignment between the measurement direction (slit) and the bar, see the discussion in \citet{Zou2019}. By contrast, the orbit-superposition method uses the entire 2D kinematic map and full LOSVD information (although $\Omega$ is mostly constrained by $v_0,s$), and the correct value of $\Omega$ is recovered even in these cases, although with larger uncertainties. 
Of course, we stress again that we used the true shape of the 3D density profile, simplifying the task of determining the best-fit parameters, while the  Tremaine-Weinberg method is model-independent. In a realistic scenario, one would try different combinations of intrinsic shape and viewing angles that are all consistent with the observed surface brightness profile, and use the kinematics to select the best combination. Our preliminary tests indicate that this procedure indeed favors the correct choice, but we leave a detailed investigation for a future study.

Figure~\ref{fig:barred_models_structure} illustrates the recovery of the internal kinematics and the orbital distribution by the Schwarzschild models. We plot the orbit circularity $\lambda_z$ -- time-averaged value of $L_z$ normalized to the maximum possible angular momentum at the given energy, $L_\mathrm{circ}(E)$, introduced in \citet{Zhu2018}. This quantity is different from the instantaneous normalized $L_z$ in a triaxial system, at least for box orbits, which have time-averaged $L_z\approx 0$, but a non-zero $L_z$ at any given time.
In the bar region, no orbits have $\lambda_z$ close to unity, because the bar rotates rather slowly in these models, and bar-trapped orbits are strongly non-circular. At larger radii, most stars are on disk orbits with $\lambda_z\simeq 1$. The gaps and bands at the transition between the bar and the disk are caused by resonances. The bottom row shows the intrinsic velocity moments in cylindrical coordinates as functions of radius. Overall, the Schwarzschild models are able to recover the orbital populations and kinematic profiles remarkably well, at least in the bar region where they are constrained by observations. We also see that regularization helps to avoid sudden and implausible variations in these quantities at large radii not covered by observations.

\section{Discussion}  \label{sec:discussion}

We present a new, publicly available implementation of the Schwarzschild orbit-superposition method for constructing equilibrium models constrained by observations. 
Its most important features are:
\begin{itemize}
\item It is applicable to systems with any shape and density profile, ranging from spherical to triaxial, including strongly flattened disks and rotating bars (in this case, the models are stationary in the rotating frame).
\item The dynamical self-consistency (if desired) is achieved by constraining the 3D density profile discretized into several types of basis elements, in particular, piecewise-linear basis functions.
\item It can deal with kinematic constraints provided in the form of classical or Gauss--Hermite velocity moments, or the full LOSVD.
\item The internal representation of the kinematic datacube uses high-accuracy 2nd or 3rd-degree B-splines.
\item Initial conditions for the orbit library are sampled randomly instead of on a regular grid, using one of several auxiliary approach such as DF inversion or Jeans equations.
\item The use of a very efficient quadratic optimization solver for determining the orbit weights makes it possible to deal with very large problems (e.g., $\mathcal O(10^5)$ orbits and $\mathcal O(10^4)$ constraints).
\item The code is highly optimized and parallelized for multi-core CPUs.
\end{itemize}

We illustrated the performance of the method on simulated datasets constructed from $N$-body or DF-based models, with parameters mimicking a Milky Way-sized galaxy at a distance of the Virgo cluster observed by a typical modern IFU. We considered several test cases: axisymmetric galaxies with a central SMBH, or a triaxial barred disk galaxy, all observed at different orientations. When using the true shape of the 3D density distribution, the code is able to recover the true values of the mass-to-light ratio and the pattern speed with small uncertainties. At the same time, with the chosen parameters of the mock datasets, we were not able to put strong constraints on the SMBH mass or on the DM halo properties.

On the other hand, we raised but did not address in detail several conceptual issues.
Most importantly, for our mock tests we assumed a known 3D shape, but in reality it needs to be inferred from the projected light distribution. This problem has no unique solution in a general case, despite the existence of methods such as MGE decomposition, which produce a solution belonging to a particular class of models. However, this class of ellipsoidally stratified profiles may not be adequate for barred disky galaxies, as illustrated in Figure~2 of \citet{IAU}.
Ideally, one would need a method for systematically exploring the range of possible shapes and orientations consistent with the observed light distribution, and determine the best-fit one by constructing a full series of Schwarzschild models for each choice of the 3D shape. Clearly, the task of exploring all possible deprojections consistent with the observed photometry can be considered independently from the task of constructing a dynamical model for each of these deprojected density profiles.

The second aspect is the intrinsic non-uniqueness of dynamical models, or more specifically, the range of possible gravitational potentials, in which the tracer population reproduces the given 3D kinematic datacube (LOSVD as a function of two sky-plane coordinates). Our tests on noiseless mock datasets demonstrate that models with a wide range of SMBH masses are able to produce perfect fits to the observed kinematic maps. However, upon closer examinations it appears that the models near the edges of this parameter space look less realistic than the models with true parameters. Namely, they have large and rapid variations in the internal structure, especially outside the range of radii covered by observed kinematics. The range of allowed potentials shrinks when increasing the spatial coverage of kinematic constraints, but it remains an open question whether this range shrink to zero in the limit of infinite coverage, or there still remains some degree of intrinsic degeneracy in the models. Furthermore, increasing the value of the regularization parameter $\lambda$ also narrows down the range of allowed potentials by eliminating the models with large variations in orbit weights. Larger values will progressively bias the solution towards the priors set by the adopted procedure for assigning initial conditions, and there is no obvious way to choose the optimal value of $\lambda$ in the noise-free case.

A third aspect, related to the previous one, is a statistically sound method for determining the confidence intervals on model parameters (in particular, $M_\bullet$) in the realistic case of noisy data. 
Our tests with mock data perturbed by several different realizations of noise indicate that the best-fit value of $M_\bullet$ often differs from the correct one by a factor of few. Moreover, the difference in $\chi^2$ between the best-fit and the true parameters ($\Delta\chi^2$) is several times larger than would have been expected for the $\chi^2$ distribution with one degree of freedom. In principle, there are no compelling reasons to expect that $\Delta\chi^2$ should satisfy that distribution, given that the models have a large number of hidden parameters (orbit weights) which are ignored in this comparison. These experiments (fitting models to many realizations of noise and examining the distribution of $\Delta\chi^2$ between the best-fit and the true parameters) may be used to calibrate the choice of threshold in $\Delta\chi^2$ for various confidence intervals, but are complicated by the noisiness of the $\chi^2$ profiles. They are also influenced by the choice of the regularization parameter $\lambda$, whose optimal choise may be guided by statistical considerations such as cross-validation.

Despite these conceptual questions, each of them probably deserving a separate study, the Schwarzschild method continues to be a powerful tool for analyzing the structure and dynamical properties of galaxies. By providing our implementation of the method to the community, we hope to reduce the entry threshold for its usage, facilitate its application to the actual observations, and catalyze research into its theoretical foundations.

\paragraph{Acknowledgements} EV acknowledges support from the European Research Council (ERC) Horizon 2020 programme under grant 308024. MV acknowledges support from HST-AR-13890.001, NSF award AST-1515001 and JWST-ERS-01364.002-A.


\appendix

\section{B-spline representation of LOSVD}  \label{sec:Bspline}

In this section we present the mathematical method for handling the LOSVD in terms of basis-set expansion using tensor-product B-splines.

A one-dimensional B-spline of degree $D$ is a piecewise polynomial defined by grid knots $x_g, g=1..G$, and can be represented as a linear combination of basis functions $e_j(x), j=1..B$ with amplitudes $f_j$: 
\begin{align}  \label{eq:BsplineExpansion}
\tilde f(x) = \sum_{j=1}^{B} f_j\,e_j(x).
\end{align}
The number of basis functions is $B=G+D-1$, each function is nonzero on at most $D+1$ consecutive intervals between knots, and has $D-1$ continuous derivatives at each knot. The case $D=0$ is equivalent to a histogram (basis elements are $\sqcap$-shaped blocks), $D=1$ -- to a linear interpolation ($\wedge$-shaped blocks spanning two grid segments), $D=3$ -- to a clamped cubic spline.

B-splines form a $B$-dimensional basis in the subset of all piecewise-continuous functions $f(x)$ on the interval $x_1..x_G$. We define the inner product of two functions $f$ and $g$ as
\begin{align}
\langle f(x),\, g(x) \rangle \equiv \int_{x_1}^{x_G} f(x) \, g(x) \,\d x ,
\end{align}
and if the second function is one of the basis elements $e_i$, we call it a projection operator
\begin{align}  \label{eq:BsplineProjection}
\mathcal P_i\{f\} \equiv \langle f(x),\, e_i(x) \rangle .
\end{align}

Any function $f(x)$ may be approximated by a B-spline $\tilde f(x)$ with the vector of amplitudes $\boldsymbol f \equiv f_i$ computed from the requirement that the projection of function $f(x)$ onto each basis element is the same as the projection of its approximated counterpart (i.e., Galerkin projection):
\begin{align}  \label{eq:BsplineMatrix}
\mathcal P_i\{ f \}= \mathcal P_i \{ \tilde f \} =
\sum_{j=1}^B M_{ij} f_j \quad \mbox{for all }i=1..B, \quad
\mbox{where the matrix }\mathsf{M}\equiv M_{ij} \equiv \langle e_i(x),\, e_j(x) \rangle.
\end{align}

When constructing the LOSVD of an orbit, we record its position and velocity at discrete moments of time $t_n, n=1..N_\mathrm{samples}$ (for simplicity, equally spaced, but this is trivially generalized). Consider, for instance, the velocity dimension (the two spatial coordinates are treated in the same way). The original, discretely sampled LOSVD is $f(v) = N_\mathrm{samples}^{-1} \sum_{n=1}^{N_\mathrm{samples}} \delta(v-v_n)$, and its projection on the $i$-th basis function is $\mathcal P_i = N_\mathrm{samples}^{-1} \sum_{n=1}^{N_\mathrm{samples}} e_i(v_n)$. The B-spline representation of the LOSVD is then given by (\ref{eq:BsplineExpansion}) with amplitudes $f_j$ found by solving the linear system $M_{ij} f_j = \mathcal P_i$.

The convolution of a function $f(x)$ with a kernel $K(x)$ is defined as
\begin{align}  \label{eq:Convolution}
f_K(x) \equiv f\ast K \equiv \int_{x_1}^{x_G} f(x')\, K(x-x')\,\d x'.
\end{align}

The B-spline approximation of the convolved function $f_K(x)$ may be constructed by the following procedure:\\
-- obtain the projections $\boldsymbol{\mathcal{P}}\{f\}$; \\
-- find the amplitudes for the B-spline approximation $\tilde f$ of the original function: $\boldsymbol{f} = \mathsf{M}^{-1} \boldsymbol{\mathcal{P}}\{f\}$; \\
-- convolve $\tilde f(x) \equiv \sum_{j=1}^B f_j\,e_j(x)$ with the kernel $K$ and obtain its projections 
\begin{align}
\mathcal{P}_i \{ \tilde f_K \} &= 
\int_{x_1}^{x_G} \Bigg[ \int_{x_1}^{x_G} \sum_{j=1}^B f_j\,e_j(x')\,K(x-x')\,\d x'\Bigg]\,e_i(x)\,\d x = \sum_{j=1}^B f_j K_{ij} ,
\end{align}
where the matrix $\mathsf{K}$ is defined as
\begin{align}  \label{eq:BsplineConvolutionMatrix}
K_{ij} &\equiv \int_{x_1}^{x_G} \d x \int_{x_1}^{x_G} \d x'\,e_i(x)\,e_j(x')\,K(x-x').
\end{align}
-- find the amplitudes of the convolution approximation:
\begin{align}  \label{eq:BsplineConvolutionAmplitudes}
\boldsymbol{f}_K = \mathsf{M}^{-1} \boldsymbol{\mathcal{P}}\{ \tilde f_K \} =
\mathsf{M}^{-1} \mathsf{K}\, \boldsymbol{f} =
\mathsf{M}^{-1} \mathsf{K}\, \mathsf{M}^{-1}\, \boldsymbol{\mathcal{P}}\{f\}.
\end{align}

Of course, the matrices $\mathsf{M}$ and $\mathsf{M}^{-1}\, \mathsf{K}\, \mathsf{M}^{-1}$ may be precomputed in advance for the given B-spline basis and kernel, so that the amplitudes $\boldsymbol{f}_K$ are obtained in one matrix--vector multiplication, requiring $B^2$ operations. Note that in the trivial case of a delta-function kernel $K(x)=\delta(x)$, the matrix $\mathsf{K} = \mathsf{M}$ and hence the convolution of the B-spline representation $\tilde f$ does not change the amplitudes, as expected.

It is also straightforward to compute integrals of $\tilde f$ over a predefined set of intervals $x_{s,\mathrm{low}}..x_{s,\mathrm{upp}}$, $s=1..S$, by expressing them as another matrix-vector multiplication operation:
\begin{align}
\mathcal I_s \equiv \int_{x_{s,\mathrm{low}}}^{x_{s,\mathrm{upp}}} \tilde f(x)\,\d x = 
\sum_{j=1}^B Q_{sj}f_j, \quad \mbox{where the matrix }
\mathsf Q \equiv Q_{sj} \equiv \int_{x_{s,\mathrm{low}}}^{x_{s,\mathrm{upp}}} e_j(x)\,\d x.
\end{align}

The LOSVD of each orbit in the model is first recorded as a projection $\mathcal P_{ijk}$ onto a 3d grid of tensor-product B-splines (two coordinates on the image plane $X,Y$ indexed by $i,j$, and the line-of-sight velocity $V$ indexed by $k$). The two indices $i,j$ are combined into a single flattened index $p$, hence the projections are represented by a matrix $\boldsymbol{\mathcal P} \equiv \mathcal P_{pk}$ with $B_X B_Y$ rows and $B_V$ columns. Each $k$-th slice of this datacube along the velocity axis is then convolved with the spatial PSF and rebinned onto the array of apertures, indexed by $s$, by integrating analytically the 2d B-spline over the area of each aperture (defined by an arbitrary non-self-intersecting polygon in the image plane). The combination of convolution and rebinning operations is described by a single matrix $\mathsf R \equiv R_{sp}$, precomputed in advance by multiplying the matrices $\mathsf M_X^{-1}, \mathsf M_Y^{-1}, \mathsf K_X, \mathsf K_Y$ and $\mathsf Q$ in the appropriate order. Finally, the projections along the velocity axis are also converted into the amplitudes $\mathsf f \equiv f_{sk}$ of B-spline expansion for $\tilde f_s(v)$ in each $s$-th aperture. The entire sequence of operations can be written as a series of matrix multiplications:
$\mathsf f = \mathsf R\, \boldsymbol{\mathcal P}\, \mathsf M_V^{-1}$,  
with the overall cost of $(B_X B_Y + B_V) B_V N_\mathrm{apertures}$ operations, which are extremely efficient on modern processors when using optimized linear algebra libraries such as \textsc{Eigen}.

\section{Accuracy tests}  \label{sec:accuracy}

\begin{figure*}
\includegraphics{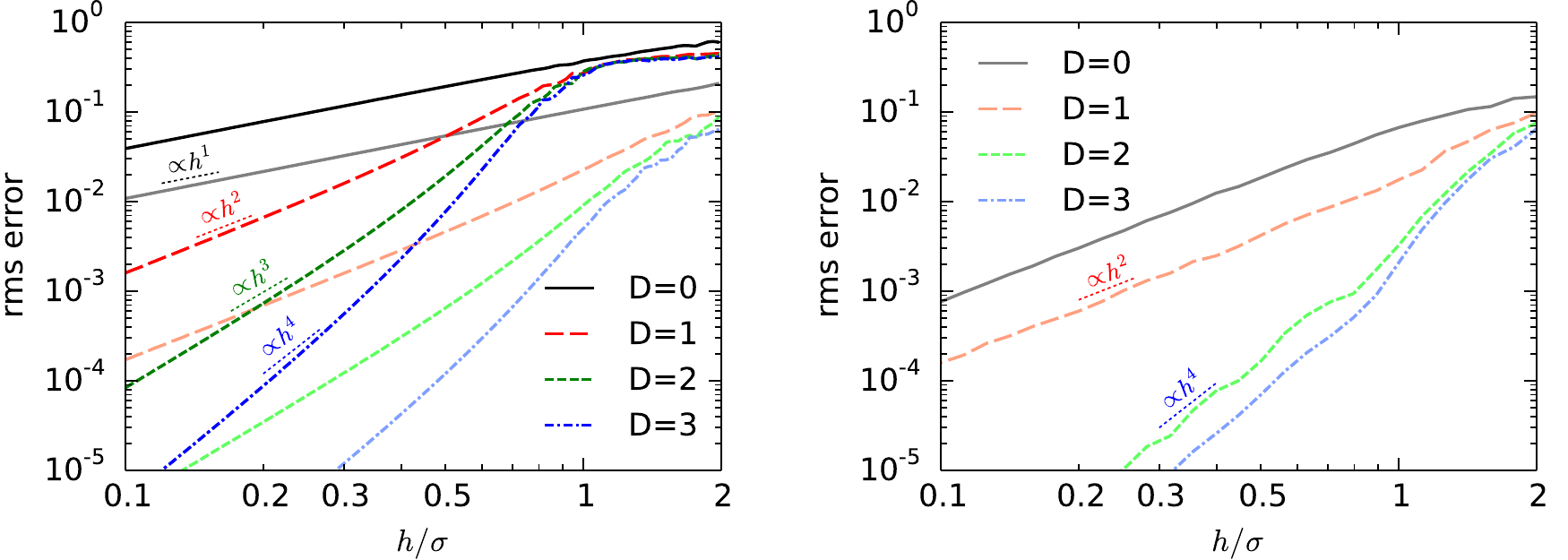}
\caption{
\textit{Left panel:} Accuracy of B-spline representation of a LOSVD.\protect\\
Shown are RMS errors $\sqrt{\int (f-\tilde f)^2\,\d v}$ in approximating a given $f(v)$ with a $D$-th degree B-spline $\tilde f(v)$, as functions of grid spacing $h$ measured in units of velocity dispersion $\sigma$. Lighter colors and lower errors are for a pure Gaussian $f(v) = \mathcal N(v)$, darker colors and higher errors -- for the 6th GH moment $f(v) = \mathcal N(v)\,\mathcal H_6(v)$, which has a higher frequency.\protect\\
With 2nd or 3rd-degree B-splines, a velocity grid spacing of $\sim (0.5-0.6)\times \sigma$ is sufficient to achieve sub-percent accuracy for any reasonable LOSVD, even if it has significant features in $h_6$; one would need to have a grid with $h\lesssim 0.1\sigma$ in order to obtain a similar accuracy with conventionally used histograms ($D=0$).\protect\\
\textit{Right panel:} Accuracy of B-spline representation and convolution of 2d functions on the image plane.\protect\\
We first construct a B-spline basis set of degree $D$ over a 2d regular grid with spacing $h$, and approximate a $\delta$ function with this basis. Then this approximation is convolved with a 2d Gaussian PSF of width $\sigma$, and re-interpolated onto the same basis. Then we compute the integral of the B-spline representation over a circular aperture with radius $\sigma$, centered on the input point, and compare it with the analytical value $1-\exp(-\frac{1}{2})\approx 0.393$.
This procedure is identical to the one used in constructing the PSF-convolved LOSVD of an orbit in a given aperture on the image plane.\protect\\
Plotted are RMS errors of this approximation, averaged over $10^3$ randomly placed points, as functions of grid spacing $h$ measured in units of PSF width $\sigma$.
With 2nd or 3rd-degree B-splines, the relative error is $\lesssim 1\%$ already at $h=\sigma$, and rapidly drops with decreasing $h$, whereas one would need $5\times$ finer grid to achieve the same accuracy in the case of histograms ($D=0$).
}  \label{fig:bspline_losvd}
\end{figure*}

In this section, we demostrate the accuracy of B-splines for representing the LOSVD.
As explained in Section~\ref{sec:kinematics}, we represent each orbit's LOSVD on a 3D grid (two image plane coordinates $X,Y$ and the line-of-sight velocity $V$) using a basis set of tensor-product B-splines of degree $D$, ranging from 0 to 3: $f(X,Y,V) = \sum_{i,j,k} A_{ijk}\;B_i(X)\, B_j(Y)\, B_k(V)$.
Each velocity slice of the datacube is then separately convolved with the spatial PSF and rebinned onto the given array of arbitrarily-shaped regions (apertures) in the image plane, in which the observed LOSVDs are measured.

The first test determines the accuracy of representation of 1d LOSVD in a given aperture in terms of B-splines. Figure~\ref{fig:bspline_losvd}, left panel, shows that a Gaussian velocity profile is approximated to better than 1\% relative accuracy with $D=2$ or $D=3$ B-splines even when the velocity-grid spacing is equal to the velocity dispersion. To resolve finer details in the LOSVD, the 6th-order GH moment is usually sufficient, and it needs roughly twice finer velocity grid. By contrast, to achieve a similar accuracy with the conventionally-used histogram representation of the LOSVD, one would need to have $5-10\times$ finer grids.
We conclude that the velocity grid should extend to $\pm 3\times$ the highest value of $\sigma$ encountered in the input data, and have a resolution $\sim 0.5\times$ the lowest value of $\sigma$. Of course, one needs to take into account that the velocities are scaled as $\Upsilon^{1/2}$ when comparing the model to the data, so the extent and resolution might need to be increased to cover the likely range of $\Upsilon$ values in the models.

The second test determines the requirements on the internal spatial grid for recording the LOSVDs of orbits. Figure~\ref{fig:bspline_losvd}, right panel, demonstrates that a grid spacing roughly equal to the PSF width is sufficient to achieve $\lesssim 1\%$ relative accuracy for $D=2$ or $D=3$ B-splines. Of course, the apertures or pixel sizes in the observed dataset need not be similar to the PSF width. If the observed data are oversampled relative to PSF, the internal grids do not need to be finer, because the rebinning procedure is geometrically exact (the contribution of each basis function of the internal grid to each observed aperture is computed analytically), and there are no features with scales smaller than PSF width in the convolved datacube. In the opposite case, when the apertures are much larger than the PSF (most commonly, in the outer parts of the galaxy), the internal grid spacing may be made comparable to the aperture widths. The B-spline grids in $x,y$ need not be uniformly spaced (for instance, they could be denser around origin) and can be tailored to the spatially-varying aperture sizes. Alternatively, separate internal datacubes may be used for two or more observational datasets with very different aperture sizes. One also needs to use a separate internal datacube for each dataset with a different PSF. In our approach, the PSF may consist of one or more circular Gaussians.

\end{document}